\documentclass{article} % For LaTeX2e
\usepackage{iclr2025_conference,times}

\usepackage{amsmath,amsfonts,bm}

\def\eqref#1{equation~\ref{#1}}
\def\1{\bm{1}}

\DeclareMathAlphabet{\mathsfit}{\encodingdefault}{\sfdefault}{m}{sl}
\SetMathAlphabet{\mathsfit}{bold}{\encodingdefault}{\sfdefault}{bx}{n}

\usepackage{hyperref}
\usepackage{url}
\usepackage{booktabs}
\usepackage{enumitem}
\usepackage{algpseudocode}
\usepackage{algorithmicx}
\usepackage{algorithm}
\usepackage{multirow}
\usepackage{graphicx}
\usepackage{subfigure}
\usepackage{caption}
\usepackage{bm,bbm}
\usepackage{colortbl}
\usepackage{wrapfig}
\usepackage{lipsum}
\usepackage{xparse}
\usepackage{array}

\definecolor{lightblue}{rgb}{0.25, 0.41, 0.88}
\definecolor{darkblue}{rgb}{0., 0.02, 0.4}
\definecolor{myorange}{rgb}{1.0, 0.62, 0.14}
\definecolor{mygreen}{rgb}{0., 0.573, 0.27}
\definecolor{myred}{rgb}{0.83, 0.078, 0.353}

\hypersetup{
  colorlinks   = true, %Colours links instead of ugly boxes
  urlcolor     = blue, %Colour for external hyperlinks
  linkcolor    = lightblue, %Colour of internal links
  citecolor   =  lightblue, %Colour of citations
}

\title{Deep Signature: Characterization of Large-Scale Molecular Dynamics}
\iclrfinalcopy
\author{Tiexin Qin\textsuperscript{1}, Mengxu Zhu\textsuperscript{1}, Chunyang Li\textsuperscript{2}, Terry Lyons\textsuperscript{3}, Hong Yan\textsuperscript{1}, Haoliang Li\textsuperscript{1}$^,$\thanks{Corresponding author.} \\
	City University of Hong Kong\textsuperscript{1} \& Chengdu Institute of Biological Products co. Ltd\textsuperscript{2} \\
 \&  University of Oxford\textsuperscript{3}\\
	\texttt{\{tiexinqin,mengxuzhu\}2-c@my.cityu.edu.hk,lichunyang@sinopharm.com}\\
    \texttt{tlyons@maths.ox.ac.uk, \{ityan,haoliang.li\}@cityu.edu.hk}
}
\newcommand{\RR}{\mathbb{R}}

\newcommand{\eg}{\textit{e}.\textit{g}.,}
\newcommand{\etc}{\textit{etc}.}

\NewDocumentCommand{\rot}{O{45} O{1em} m}{\makebox[#2][l]{\rotatebox{#1}{#3}}}%

\newtheorem{theorem}{Theorem}

\begin{document}

\maketitle

\begin{abstract}
Understanding protein dynamics are essential for deciphering protein functional mechanisms and developing molecular therapies. However, the complex high-dimensional dynamics and interatomic interactions of biological processes pose significant challenge for existing computational techniques. In this paper, we approach this problem for the first time by introducing Deep Signature, a novel computationally tractable framework that characterizes complex dynamics and interatomic interactions based on their evolving trajectories. Specifically, our approach incorporates soft spectral clustering that locally aggregates cooperative dynamics to reduce the size of the system, as well as signature transform that collects iterated integrals to provide a global characterization of the non-smooth interactive dynamics. Theoretical analysis demonstrates that Deep Signature exhibits several desirable properties, including invariance to translation, near invariance to rotation, equivariance to permutation of atomic coordinates, and invariance under time reparameterization. Furthermore, experimental results on three benchmarks of biological processes verify that our approach can achieve superior performance compared to baseline methods. Our code is available at \href{https://github.com/WonderSeven/Deep-Signature}{https://github.com/WonderSeven/Deep-Signature}.
\end{abstract}

% ======================================================================= 
\section{Introduction}
\label{sec:intro}
Biological processes are fundamentally driven by the dynamical changes of macromolecules, particularly proteins and enzymes, within their respective functional conformation spaces. Typical examples of such processes include protein–ligand binding, molecule transport and enzymatic reactions, and modern computational biologists investigate their underlying functional mechanisms by molecular dynamics (MD) simulations~\citep{Dror2012ARB, Lewandowski2015Science}. Built upon density functional theory~\citep{Car1985PRL}, MD has demonstrated remarkable capability in providing accurate atomic trajectories in three-dimensional (3D) conformational space and consist agreement with experimental observations~\citep{Frenkel2023Understanding}. 

The computational analysis of MD data has been a subject of extensive research for decades, with the goal of characterizing systems from trajectory information. However, due to the main challenge posed by intricate interatomic interactions over large-scale systems across inconsistent timescales, many existing works resort to oversimplified setups that incorporate biophysical priors to analyze certain aspects of dynamics such as protein fluctuations, relaxation time, stability, and state transitions~\citep{Law2017JACS,Qiu2023JCTC}. More recently, empowered by the parallel processing ability of GPUs, machine learning especially deep learning sheds new light on this field as it can discretize macromolecules as particles distributed in a 3D voxel grid and automatically learn their relations in a data-driven fashion~\citep{Li2020PLoS,Rogers2023Growing}. Parallel to this, surface modeling-based approaches have emerged, firstly utilizing mathematical models to restore protein surfaces and then applying deep learning to analyze the chemical and geometrical features of surface regions around binding sites~\citep{Gainza2020NM,Zhu2021BMC}. Despite the great potential of these approaches in automatic drug discovery, their computational complexity would increase linearly with the number of time stamps when processing MD data, struggling with application to long-time simulations. Besides, these methods commonly build upon coarse grained dynamics for accelerating computation. Nevertheless, selecting an optimal coarse graining mapping strategy that effectively simplifies the representation of the system while preserving essential features remains an open research problem~\citep{Jin2022JCTC,Majewski2023NC}.

Another limitation of current MD analysis methods is the deficient utilization of structural bioinformatics for the largely increased difficulty in handling high-order interatomic interactions during dynamic processes. However, such structural bioinformatics, manifested in various covalent and non-covalent bonds, plays a pivotal role in molecular design for its capability of propagating local perturbations to facilitate conformational dynamics and alter biological function~\citep{Tsai2009MB,Otten2018NC}. An illustrative example would be dihydrofolate reductase, which has been widely studied as important antitumor and antibacterial targets for treating tuberculosis and malaria. There exist four common mutations that confer drug resistance to antibiotics, proceeding in a stepwise fashion. Among them, the P21L mutation acts in a dynamical loop region associated with long range structural vibrations of the protein backbone, rather than directly on the active sites as other mutations~\citep{Toprak2012NG}. Therefore, ignoring such interatomic interactive dynamics facilitated by molecular structure for a critical protein and counting the effects of active sites solely can result in biased assessments of designed drugs. Nonetheless, since the integration of structural bioinformatics into MD analysis would further introduce at least quadratic complexity with system size, existing works have not yet investigated this crucial aspect, highlighting a critical gap in our ability to comprehensively analyze and predict molecular behavior in drug design and resistance studies.

To this end, we aim to develop a computationally efficient framework that incorporates the structural bioinformatics with coarse graining mapping for automatically analyzing protein trajectory dynamics. In particular, we first introduce a graph clustering module that learns to extract coarse grained dynamics by approximating soft spectral clustering. With the clustering assignment function implemented by a graph neural network and parameters learned automatically, we circumvent the need for manual selection of coarse graining mapping. Subsequently, we introduce a path signature transform module served as a feature extractor to characterize the interatomic interactive dynamics after coarse graining. \textcolor{black}{Path signature is a mathematically principled concept that utilizes iterated integrals to describe geometric rough paths in a compact yet rich manner~\citep{Terry2014ICM}, thus suitable for our tasks where molecular trajectories are highly sampled and non-smooth.} After attaching with a task-specific differentiable classifier or regressor, we devise an end-to-end framework, named Deep Signature, for efficiently characterizing the complex protein dynamics. Notably, due to the existence of considerable random fluctuation in simulated trajectories, ideal features ought to maintain symmetry respecting certain geometric transformations. We provide theoretical analysis that our extracted features exhibit invariance to translation, near invariance to rotation, equivariance to permutation of atomic coordinates, and invariance under time reparametrization of paths. Finally, we target our task on predicting functional properties of proteins from MD data, a fundamental task for developing novel drug therapies. We consider three benchmarks including gene regulatory dynamics, epidermal growth factor receptor (EGFR) mutation dynamics, and G protein-coupled receptors (GPCR) dynamics for performance evaluation. The contributions of our paper are as follows:
% \vspace{-8pt}
\begin{itemize}%[align=left,leftmargin=*,widest={10}]
    \item We develop Deep signature, the first computationally efficient framework that characterizes the complex interatomic interactive dynamics of large-scale molecules.
    % \vspace{-3pt}
    \item We theoretically demonstrate that our approach preserves symmetry under several geometrical transforms of atomic coordinates in 3D conformational space. Additionally, our method remains invariant under time reparameterization.
    % \vspace{-3pt}
    \item We provide empirical results to show that our Deep Signature model achieves superior performance compared to other baseline methods on gene regulatory dynamics, EGFR mutation dynamics and GPCR dynamics.
\end{itemize}
\vspace{-0.11cm}

% ======================================================================= 
\section{Related Works}
\label{sec:related_work}
\vspace{-0.16cm}
\noindent\textbf{Molecular Representation.}~Encoding essential structural characteristics and biochemical properties into molecular representations is a long-standing research field in molecular biology, with wide applications in various drug discovery processes including virtual screening, similarity-based compound searches, target molecule ranking, drug toxicity prediction, and \etc~\citep{Li2024Bio}. One of the most widely used categories of molecular representation is two-dimensional fingerprints that extract the substructure, topological routes and circles solely from molecular connection tables. Owing to their ease of generation and usage, these 2D fingerprints are still extensively utilized as input for machine learning algorithms in modern drug discovery applications~\citep{Gao2020PCCP}. In recent years, there has been a surge in the development of 3D structure-based molecular representations for their much more fine-grained characterization of interatomic interplay in 3D conformational space. Surface modeling-based approaches leverage mathematical models to restore protein surfaces and encode geometrical and chemical features of surface regions around binding sites into representations, exhibiting great potential in automatic drug discovery~\citep{Gainza2020MaSIF,Zhu2021BMC}. In addition to surface-based representations, significant efforts have been devoted to learning accurate deep learning force fields for accelerating MD simulation~\citep{Schutt2017NIPS,Batatia2022NIPS,Batzner2022NC}. While these methods can be adapted for molecular property prediction, they are limited to generating representation for a static frame, thus not suitable for our tasks where interframe interactions along timescales are crucial for understanding protein function.

The current investigation into characterizing molecular dynamics, especially interatomic interaction dynamics, remains limited, with only a few studies close to ours. Among them, \cite{Endo2019Nano}, \cite{Yasuda2022CB} and \cite{Mustali2023RSC} are unsupervised methods that build local dynamics ensembles for pre-specified atoms and inspect each atom's contribution independently. \cite{Li2022JCIM} converts MD conformations into images and applies convolutional neural networks to identify diverse active states. \textcolor{black}{\cite{Sun2023DSR} models protein dynamics by representing protein surfaces using implicit neural networks without requiring explicit surface representations.} Nevertheless, these approaches cannot capture subtle interatomic interactions along atomic pathways for dissecting protein function, nor is the complexity of their representations independent of time stamps.

\noindent\textbf{Coarse Graining (CG).}~CG is a widely adopted technique with the objective of preserving the crucial characteristics and dynamics inherent to a molecular system. This is achieved by grouping sets of atoms into CG beads, thereby enabling high-throughput MD simulations over larger time and length scales. Existing CG methods can be broadly categorized into two types: chemical and physical intuition-based approaches and machine learning-based approaches. The methods of first type construct the CG mapping by incorporating various biochemical properties derived from expert knowledge, for example, mapping each elaborately constructed cluster of four heavy (nonhydrogen) atoms into a single CG bead~\citep{Marrink2013CSR} or simply assigning one CG bead centered at the $\alpha$-carbon for each amino acid~\citep{Ingolfsson2014CMS}. Machine learning-based methods can rapidly learn accurate potential energy functions for reduced structures of MD by training on large databases. Recent advancements in this area include multiscale coarse graining that optimizes to maximize a variational force matching score~\citep{Wang2019NPJ}, relative entropy minimization~\citep{Thaler2022JCP}, and spectral graph approaches that account for structural typologies of proteins~\citep{Webb2018JCTC,Li2020CS}. However, their transferability to unseen molecules remains suspicious, and the representation capability for complex macromolecular systems without increasing dimension and complexity is still underexplored~\citep{Khot2019JCP}.

% ======================================================================= 
\section{Methodology}
\label{sec:methodology}

\begin{figure*}[!tbp]
% \vspace{-0.1cm}
\centering
\includegraphics[width=0.82\textwidth]{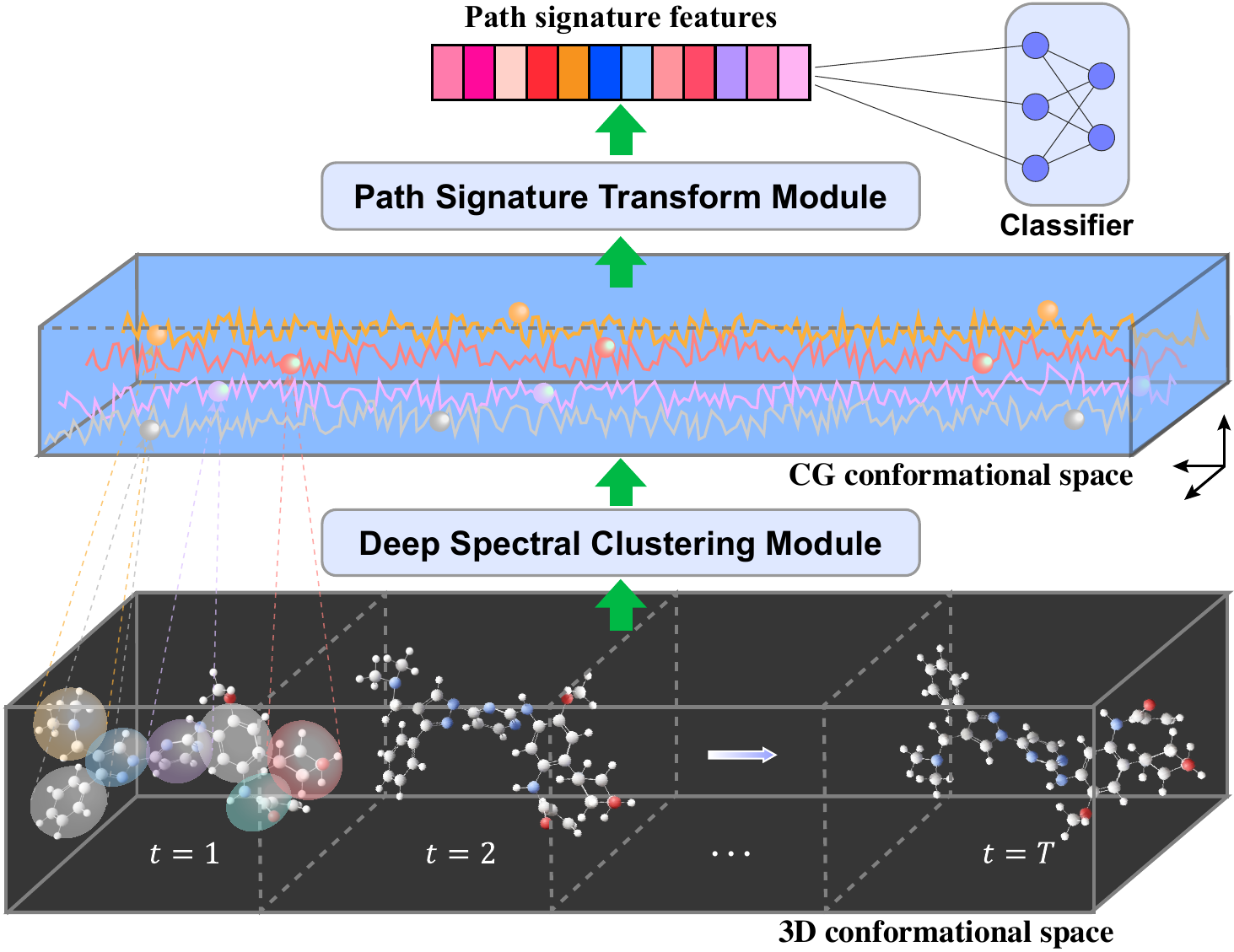}
\vspace{-0.2cm}
\caption{An overview of our proposed Deep Signature method.}
\label{fig:framework}
% \vspace{-0.2cm}
\end{figure*}

% -----------------------------------------------------------------------
\subsection{Problem Formulation}
Consider various molecular systems $\smash{S^{(k)}}$ that are distinct in their molecular behaviors. An MD simulation trajectory on $\smash{S^{(k)}}$ provides the trajectories for all \scalebox{0.95}{$\smash{N_k}$} atoms constituting the molecules, and can be represented as a sequence of snapshots \scalebox{0.95}{$\mathbf{X}_{1:T_k}^{(k)}=\{\mathbf{X}^{(k)}_1,\mathbf{X}^{(k)}_2,\dots,\mathbf{X}^{(k)}_{T_k}\}$}. Here, \scalebox{0.95}{$\mathbf{X}^{(k)}_{t}\in\RR^{N_k\times 3}$} indicates the atomic positions in 3D conformational space at time step $t$ for $t\in\{1,\dots,T_k\}$ and $\smash{T_k}$ is the total number of frames. To describe the structure of molecules within $\smash{S^{(k)}}$, we define a molecular graph $\smash{\mathcal{G}^{(k)}=\{\mathcal{E}^{(k)},\mathcal{V}^{(k)}\}}$, $\smash{|\mathcal{V}^{(k)}|=N_k}$, where $\smash{\mathcal{V}^{(k)}}$ is the node set corresponding to the atoms and $\smash{\mathcal{E}^{(k)}}$ is the edge set corresponding to chemical bonds. The adjacency matrix of $\smash{\mathcal{G}^{(k)}}$ is represented by $\smash{\mathbf{A}^{(k)} \in \mathbb{R}^{N_k \times N_k}}$, with \scalebox{0.95}{$\smash{\mathbf{A}^{(k)}_{i,j}}=1$} if $\smash{v_i, v_j \in \mathcal{V}^{(k)}}$ and $\smash{(v_i, v_j) \in \mathcal{E}^{(k)}}$. For the molecular property prediction task, we have access to MD trajectories from $K$ molecular systems, each endowed with property labels denoted as \scalebox{0.95}{$\smash{\{(\mathbf{X}_{1:T_k}^{(k)}, y^{(k)})\}_{k=1}^K}$}. The objective is to train algorithms to accurately predict the property label when provided with an previously unseen MD trajectory \scalebox{0.95}{$\smash{\mathbf{X}_{1:T_k}}$}.

% -----------------------------------------------------------------------
\subsection{Deep Signature}
\label{sec:method}
Our proposed method, referred to as Deep Signature, consists of a deep spectral clustering module that uses GNNs to extract coarse grained dynamics from raw MD trajectories, a path signature transform module that collects iterated integrals to characterize interatomic interactions along pathways, and a classifier to enable property prediction. The overall architecture is illustrated in Fig.~\ref{fig:framework}.

\noindent\textbf{Deep spectral clustering with GNNs.}~Given the MD trajectory $\smash{\mathbf{X}_{1:T}}$ and molecular graph $\mathcal{G}$ for a molecular system $S$, we start with extracting the reduced trajectory \scalebox{0.95}{$\smash{\tilde{\mathbf{X}}_{1:T}}$} by coarse gaining $\smash{\mathbf{X}_{1:T}}$ using deep spectral clustering. Specifically, we first obtain node representations via GNN layers as
\begin{equation}
\label{eq:gcn}
    \mathbf{H}^{l} = \sigma(\tilde{\mathbf{D}}^{-1/2}\tilde{\mathbf{A}}\tilde{\mathbf{D}}^{-1/2}\mathbf{H}^{l-1} \mathbf{W}^{l-1}_{\mathrm{GNN}}),
\end{equation}
where $\smash{\mathbf{H}^{l}}$ denotes the node feature matrix at the $l$-th layer, $\smash{\mathbf{H}^{0}=\mathbf{X}_{1:T}}$, $\smash{\tilde{\mathbf{A}}=\mathbf{A}+\mathbf{I}}$ is the adjacency matrix $\mathbf{A}$ plus the identity matrix $\mathbf{I}$, $ \smash{\tilde{\mathbf{D}}}$ is the degree matrix of $\smash{\tilde{\mathbf{A}}}$, \scalebox{0.95}{$\smash{\mathbf{W}^{l-1}_{\mathrm{GNN}}}$} are the learnable parameters of GNNs, and $\sigma$ is a nonlinear activation function. After then, we compute the cluster assignment matrix $\mathbf{Q}$ for the nodes using a multi-layer perceptron (MLP) with softmax on the output layer 
\vspace{-2pt}
\begin{equation}
\label{eq:assig_func}
\mathbf{Q} = \textit{Softmax}(\mathbf{W}_{\mathrm{MLP}} \mathbf{H}^l + \mathbf{b}),
\end{equation}
where $\mathbf{W}_{\mathrm{MLP}}$ and $\mathbf{b}$ are trainable parameters of the MLP. For the assignment matrix $\mathbf{Q}\in\RR^{N\times M}$, where $M \ll N$ specifies the number of clusters, each row of it represents the node’s probability of belonging to a particular cluster. We employ the normalized-cut relaxation~\citep{Bianchi2020ICML} as our clustering objective for minimization
\vspace{-2pt}
\begin{equation}
\label{eq:loss_graph}
    \mathcal{L}_u = -\frac{Tr(\mathbf{Q}^T\tilde{\mathbf{A}}\mathbf{Q})}{Tr(\mathbf{Q}^T\tilde{\mathbf{D}}\mathbf{Q})} + \left\Vert\frac{\mathbf{Q}^T\mathbf{Q}}{||\mathbf{Q}^T\mathbf{Q}||_F} - \frac{\mathbf{I}_M}{\sqrt{M}}\right\Vert_F,
\end{equation} 
where the first term promotes strongly connected components to be clustered together, while the second term encourages the cluster assignments to be orthogonal and have similar sizes.

Upon leveraging $\mathbf{Q}$ from Eq.~(\ref{eq:assig_func}) for clustering, the corresponding reduced feature embedding matrix $\smash{\mathbf{H}'}$ and adjacency matrix $\smash{\mathbf{A}'}$ can be derived as follows
\vspace{-2pt}
\begin{equation}
   \mathbf{H}'=\mathbf{Q}^T\mathbf{H};\quad \mathbf{A}'=\mathbf{Q}^T\tilde{\mathbf{A}}\mathbf{Q}.
\end{equation}
Since our model takes the sequence $\smash{\mathbf{X}_{1:T}}$ as input, $\smash{\mathbf{H}'}$ inherently maintains the temporal order in the form of $\smash{\mathbf{H}'_{1:T}}$. We utilize another MLP with the parameters $\smash{\mathbf{W}'_{\mathrm{MLP}}}$ and $\mathbf{b}$ to map $\smash{\mathbf{H}'_{1:T}}$ back into a reduced conformational space with the resulting dynamics and ground truth dynamics expressed as
% \vspace{-2pt}
\begin{equation}
\label{eq:coarse_grain}
    \tilde{\mathbf{X}}_{1:T} = \mathbf{W}'_{\mathrm{MLP}} \mathbf{H}'_{1:T} + \mathbf{b}';\quad \tilde{\mathbf{X}}_{1:T}^{\mathrm{pool}} = \mathbf{Q}^T \mathbf{X}_{1:T}.
\end{equation}
To ensure the fidelity of the coarse grained dynamics towards the original high-dimensional system, we further introduce a temporal consistency constraint, defined through a mean absolute error loss function with the form
% \vspace{-3pt}
\begin{equation}
\label{eq:loss_temporal}
    \mathcal{L}_t = \frac{1}{T} \sum_{i=1}^{T}\big|\tilde{\mathbf{X}}_i - \tilde{\mathbf{X}}_i^{\mathrm{pool}}\big|.
\end{equation}

% To provide a more precise definition, 
\noindent\textbf{Path signature transform.}~We now adopt the path signature method to characterize the interatomic temporal interactions among the coarse grained dynamics $\tilde{\mathbf{X}}_{1:T}\in\RR^{T\times 3M}$. The basic idea of path signature is that, for a multidimensional continuous path, we can construct an ordered set consisting of all possible path integrals and combinations involving the path integrals among various individual dimensions as a comprehensive representation for this path~\citep{Terry2014ICM}. Striving for a more precise definition, consider our coarse grained trajectory $\tilde{\mathbf{X}}_{1:T}$ with $(\tilde{\mathbf{X}}^1_t, \tilde{\mathbf{X}}^2_t, \dots, \tilde{\mathbf{X}}^{3M}_t)$ for $t\in\{1,\dots,T\}$, let us define $\widehat{\mathbf{X}}:[1,T] \rightarrow \RR^{3M}$ as a piecewise linear interpolation of $\tilde{\mathbf{X}}_{1:T}$ such that $\widehat{\mathbf{X}}_t=\tilde{\mathbf{X}}_{t}$ for any $t\in\{1,\dots,T\}$, and a sub-time interval $[r_i, r_{i+1}]$ corresponding to a time partition of $[1, T]$ with $\smash{1=r_1<r_2<\cdots<r_{\tau}=T}$. The depth-$D$ signature transform of $\smash{\tilde{\mathbf{X}}}$ over the interval $\smash{[r_i, r_{i+1}]}$ is the vector defined as
\begin{equation}
\label{eq:signature}
    \mathrm{Sig}^D_{r_i, r_{i+1}}(\tilde{\mathbf{X}})=\left(1, \left\{S(\tilde{\mathbf{X}})_{r_i, r_{i+1}}^j\right\}_{j=1}^{3M},\ldots,\left\{S(\tilde{\mathbf{X}})_{r_i, r_{i+1}}^{j_1,\ldots,j_d}\right\}_{j_1, \ldots,j_d=1}^{3M} \right),
\end{equation}
where for any $\smash{(j_1,\ldots,j_d)\in\{1,\dots,3M\}^D}$,
\begin{equation}
\label{eq:sig_integral}
    S(\tilde{\mathbf{X}})_{r_i, r_{i+1}}^{j_1,\ldots,j_d}=\underset{1<t_1<\ldots <t_d<T}{\int\dots\int} d \widehat{\mathbf{X}}_{t_1}^{j_1} \ldots d \widehat{\mathbf{X}}_{t_d}^{j_d}.
\end{equation}
After that, we take the logarithm of the signature transform features presented in Eq.~(\ref{eq:signature}) to eliminate redundant elements according to the shuffle product identity~\citep{Terry2007Springer} and acquire some minimal collection of the stream over a time interval. Specifically, given the vector space described by depth-$D$ signature in formal power series form as
\begin{equation}
\label{eq:sig_power_series}
    \mathrm{Sig}^D_{r_i, r_{i+1}}(\tilde{\mathbf{X}}) = \sum^{D}_{d=0}\sum_{j_1,\ldots,j_d\in\{1,\dots,3M\}} S(\tilde{\mathbf{X}})_{1, T}^{j_1,\ldots,j_d} e_{j_1}\dots e_{j_d}, 
\end{equation}
The depth-$D$ log-signature transform corresponds to taking the formal logarithm of Eq.~(\ref{eq:sig_power_series}), which can be expressed as
\begin{equation}
\label{eq:log_sig}
    \mathrm{LogSig}^D_{r_i, r_{i+1}}(\tilde{\mathbf{X}}) = \sum_{n=1}^{3M}S(\tilde{\mathbf{X}})^n_{r_i, r_{i+1}}e_n + \sum_{1 \leq n < m \leq 3M} \frac{1}{2}(S(\tilde{\mathbf{X}})^{n,m}_{r_i, r_{i+1}} - S(\tilde{\mathbf{X}})^{m,n}_{r_i, r_{i+1}})[e_n,e_m] + \dots
\end{equation}
where $[e_n,e_m]$ is the Lie bracket~\citep{Reizenstein2017arXiv} defined by $[e_n,e_m]=e_n\times e_m - e_m\times e_n$. Generally, log-signature possesses a lower dimension compared to the original signature feature while carrying exactly the same information. \textcolor{black}{Through this transform, the interatomic dynamic correlations in 3D space are embedded as geometric areas into features. The resulting signature sequence, denoted as $(\mathrm{LogSig}^D_{r_1, r_2}(\tilde{\mathbf{X}}), \mathrm{LogSig}^D_{r_2, r_3}(\tilde{\mathbf{X}}),\dots)$, can be regarded as a discretization of dynamic data. In this work, we further utilize LSTM to tackle the nonlinear interactions between them.}
% decouple the length of representation with regards to the timescale

\noindent\textbf{Classifier.}~Finally, we introduce a classifier $f$ implemented as a two-layer MLP which outputs the predicted label $\hat{y}=f(\mathrm{LogSig}^d_{1,T}(\tilde{\mathbf{X}}))$ for molecular property prediction. To establish the classification objective function, we leverage cross-entropy loss defined as follows
\begin{equation}
\label{eq:loss_class}
\mathcal{L}_c = -y \log(\hat{y}) - (1 - y) \log(1 - \hat{y}),
\end{equation}
where $y$ is the ground truth label for the system and $\hat{y}$ is the predicted label generated by our model. 

By combining the loss terms from Eqs.~(\ref{eq:loss_graph}), (\ref{eq:loss_temporal}) and (\ref{eq:loss_class}), we arrive at the overall loss function for training our Deep Signature model as
\begin{equation}
\label{eq:loss_total}
    \mathcal{L} = \lambda_1\mathcal{L}_u + \lambda_2\mathcal{L}_t + \lambda_3\mathcal{L}_c.
\end{equation}
Here, $\lambda_1$, $\lambda_2$ and $\lambda_3$ are scaling parameters to balance the contributions of these three loss terms.

\textcolor{black}{It is worth noting that, recent research has delved into the integration of path signatures with GNNs for spatial-temporal modeling, and marked significant strides in traffic forecasting scenarios~\citep{Choi2023TIST,Wang2024MFG}. These approaches utilize path signatures to process temporal information for each node individually at a low level, while employing GNNs empowered neural differential equation at a higher level to ensure the preservation of spatial topological consistency during dynamic forecasting. Nonetheless, they fall short in characterizing the temporal interactions among different nodes, making them unsuitable for molecular property prediction tasks where the interactions among dominant kinetic pathways play an essential role in protein function~\citep{Araki2021NC,Mustali2023RSC}.}
% \textcolor{orange}{Ques: Why not follow spatial-temporal prediction models use GNN-based differential equation? Ans: identify critical paths \& interpretation.}
% 保持空间相关性在属性预测这种high-level的任务中起到的作用并没有traffice flow prediction那么重要

% -----------------------------------------------------------------------
\subsection{Equivariance \& invariance of log-signature}
\label{sec:equivariance_invariance}
Physical properties often exhibit well-defined symmetry characteristics, and integrating such equivariance into our learned feature space can enhance interpretability and mitigate learning difficulty. In the subsequent analysis, we rigorously examine the equivariance of our deep signature in relation to various geometric transformations including translation, rotation, and permutation on atomic coordinates, as well as reparameterization over time. The complete proofs are provided in Appendix~\ref{app:property_proof}.
% Semi-Supervised Hierarchical Graph Classification
% on SE(3) equivariance

\noindent\textbf{Translation invariance.}~
The coarse-grained dynamics acquired by a linear mapping in Eq.~(\ref{eq:coarse_grain}) demonstrate that $\smash{\tilde{\mathbf{X}}_{1:T}}$ maintain equivariance with respect to translations on input trajectory $\smash{\mathbf{X}_{1:T}}$. Besides, since the path signature is composed of iterated integrals, it inherits the property of translation invariance. As a result, log-signature features are translation-invariant for the input trajectory.

\noindent\textbf{Rotation invariance.}~When a rotation is applied to the 3D conformational space, an equivalent rotation exists in the coarse grained conformational space, 
%indicating that the coarse grained dynamics $\smash{\tilde{\mathbf{X}}_{1:T}}$ are equivariant with respect to rotations of the input trajectory $\smash{\mathbf{X}_{1:T}}$. 
indicating the equivariance of coarse grained dynamics $\smash{\tilde{\mathbf{X}}_{1:T}}$ to rotations of the input trajectory $\smash{\mathbf{X}_{1:T}}$.
Furthermore, while research on rotation invariance of path signatures for higher depth and multi-dimensional paths is still limited~\citep{Diehl2013arXiv}, we demonstrate that the majority of elements constituting our Deep Signature features are rotation-invariant, particularly in large-scale systems like the molecular systems studied in this work.

\noindent\textbf{Permutation equivariance.}~GCNs that aggregate contributions from neighboring atoms are invariant to permutations of those atoms, which implies that our coarse-grained dynamics are equivariant to permutations of atom indices. Additionally, the multi-indices used to index the iterated integrals in a signature are sorted in ascending order during implementation, ensuring that our signature features remain equivariant to permutations. This property is essential for facilitating the interpretability of our method, as it allows us to trace the dominant kinetic pathways that contribute to protein function.

\noindent\textbf{Time-reparametrization invariance.}~Path signature possesses a powerful property that it remains invariant under time-reparameterization of the underlying stream~\citep{Terry2014ICM}. This property substantially decreases the complexity of certain challenges by eliminating dependence on the sampling rate while preserving all other relevant information within the stream. Furthermore, it enhances the robustness of our method against deviations in molecular dynamics occurring over time scales.

% -----------------------------------------------------------------------
\subsection{Implementation}
\begin{wrapfigure}{r}{0.55\textwidth}
\centering
\vspace{-0.45cm}
\includegraphics[width=.98\linewidth]{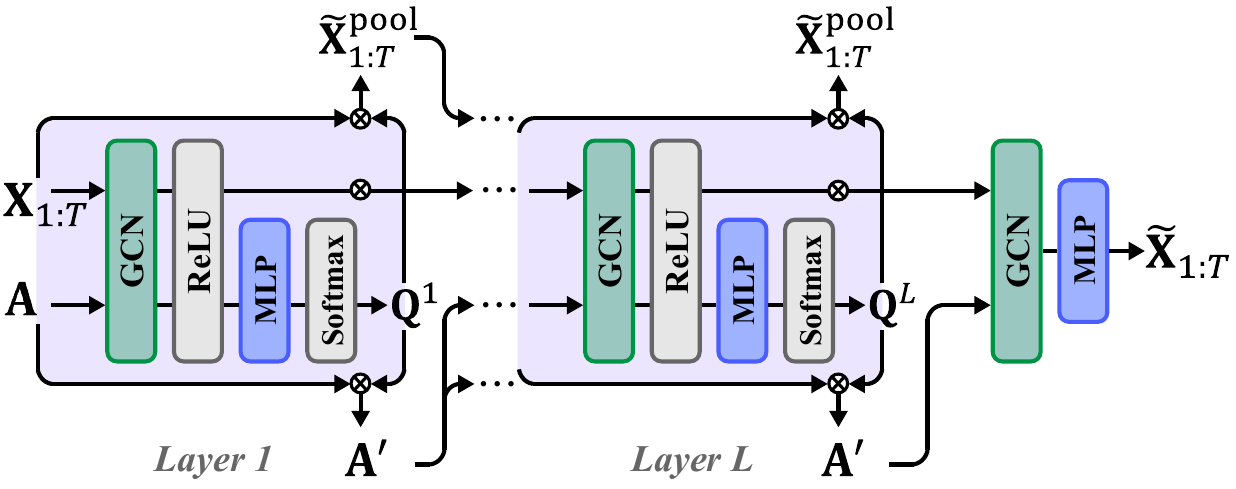}
\vspace{-0.3cm}
\caption{The architecture of deep spectral clustering module.}
\label{fig:graph_module}
\vspace{-0.4cm}
\end{wrapfigure}
\textbf{Architecture.}~Our introduced framework consists of three fundamental modules: a deep spectral clustering module, a path signature transform module and a classifier. We adopt a hierarchical pooling architecture to implement the deep spectral clustering module, which is schematically depicted in Fig.~\ref{fig:graph_module}. As shown, it contains a stack of $L$ graph pooling layers with each layer consisting of a GCN layer for obtaining node embeddings and an MLP layer for cluster assignment, such that it would gradually coarsen the dynamics from the atomic level. The last GCN layer and MLP take the node feature matrix and adjacency matrix processed by the $L$-th graph pooling layer as input and output the final coarse grained dynamics $\tilde{\mathbf{X}}_{1:T}$. 

\begin{wrapfigure}{r}{0.4\textwidth}
\centering
% \vspace{-0.45cm}
\includegraphics[width=.98\linewidth]{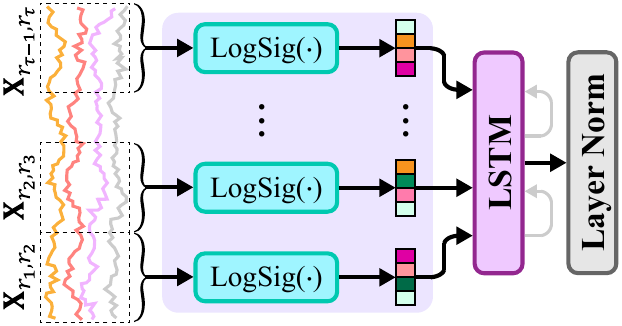}
\vspace{-0.3cm}
\caption{The architecture of path signature transform module.}
\label{fig:sig_module}
\vspace{-0.6cm}
\end{wrapfigure}
The implementation of the path signature transform module is illustrated in Fig.~\ref{fig:sig_module}. It takes $\tilde{\mathbf{X}}_{1:T}$ as input, which is first partitioned into equally spaced intervals. The resulting segments are subsequently processed through the $\mathrm{LogSig}(\cdot)$ function to extract log-signature features.
This is implemented based on \texttt{Signatory}\footnote{\url{https://github.com/patrick-kidger/signatory}}, a Python package that facilitates differentiable computations of the signature and log-signature transforms on both CPU and GPU. The resulting signature sequence is then fed into an LSTM layer to capture their interactions and followed by layer normalization~\citep{Ba2016LN} to standardize the feature values. For the classifier, we utilize a two-layer MLP together with a sigmoid activation function to enable property prediction. It is noteworthy that the overall architecture is fully differentiable, thus we can leverage off-the-shelf optimization techniques to train our model in an end-to-end fashion.

\noindent\textbf{Cross validation and independent test.}~We employ a five-fold cross-validation strategy for training our model. In contrast to conventional random division, we take the temporal nature of the trajectory data into consideration for data partition. Specifically, each trajectory is divided into 5 groups with the same time interval according to its temporal order. Subsequently, we further partition the data within each group into five folds. The validation set is constructed by selecting one fold from each group and is employed for model selection, while the remaining four folds within each group are gathered to form the training set. This process is repeated five times sequentially, resulting in the creation of the five-fold cross-validation dataset. Moreover, for each running, we evaluate the prediction accuracy of our method on an independent unseen test set and report the averaged results.

% ======================================================================= 
\section{Experiments}
\label{sec:experiments}
In this section, we conduct experiments on representative molecular dynamic systems to evaluate the effectiveness of our proposed method. We begin with a synthetic dataset that reports gene regulatory dynamics~\citep{Gao2016Nature}. We then assess the performance on two large-scale MD simulation datasets, including epidermal growth factor receptor (EGFR) mutant dynamics~\citep{Zhu2021BMC} and G protein-coupled receptors (GPCR) dynamics~\citep{Rodriguez2020GPCR}. More details on dataset construction can be found in Appendix~\ref{app:exp_setup}.

% -----------------------------------------------------------------------
\subsection{Experiments on Gene regulatory dynamics}
\begin{figure}[tbp]
    \begin{minipage}{.38\linewidth}
    \vspace{-0.4cm}
    \centering
    % \fontsize{8.3}{9.3}\selectfont
    \captionof{table}{Comparisons of classification performance on gene regulatory dynamics. Results are averaged over 5 runs.}
    \small{
    \begin{tabular}{p{58pt}<{\raggedright}p{29pt}<{\centering}p{29pt}<{\centering}}
    \toprule[0.7pt]
    \rule{0pt}{2ex} \textbf{Method} & \textbf{Accuracy} & \textbf{Recall} \\
    \hline  
    \rule{0pt}{2ex} Head            & 55.28\scalebox{0.65}{$\pm10.56$} & 30.65\scalebox{0.65}{$\pm16.23$} \\
    \rule{0pt}{2ex} Tail            & 55.36\scalebox{0.65}{$\pm10.72$} & 30.71\scalebox{0.65}{$\pm16.36$} \\ 
    \rule{0pt}{2ex} Head \& Tail    & 67.92\scalebox{0.65}{$\pm15.17$} & 35.80\scalebox{0.65}{$\pm9.27$} \\ 
    \rule{0pt}{2ex} GraphLSTM       & 96.40\scalebox{0.65}{$\pm0.89$}  & 93.14\scalebox{0.65}{$\pm1.20$} \\
    \rule{0pt}{2ex} GraphTrans      & 86.00\scalebox{0.65}{$\pm6.67$}  & 79.36\scalebox{0.65}{$\pm16.37$} \\
    \rule{0pt}{2ex} Deep Signature  & 99.12\scalebox{0.65}{$\pm0.82$}  & 98.65\scalebox{0.65}{$\pm0.23$} \\ 
    \bottomrule[0.7pt]
    \end{tabular}
    }
    \label{tab:exp_gene}
    \end{minipage}\qquad\hfill
    \begin{minipage}{.58\linewidth}
    \vspace{-0.2cm}
    \subfigcapskip=-5pt
    \subfigure[Different timescales]{
    \includegraphics[width=.47\linewidth]{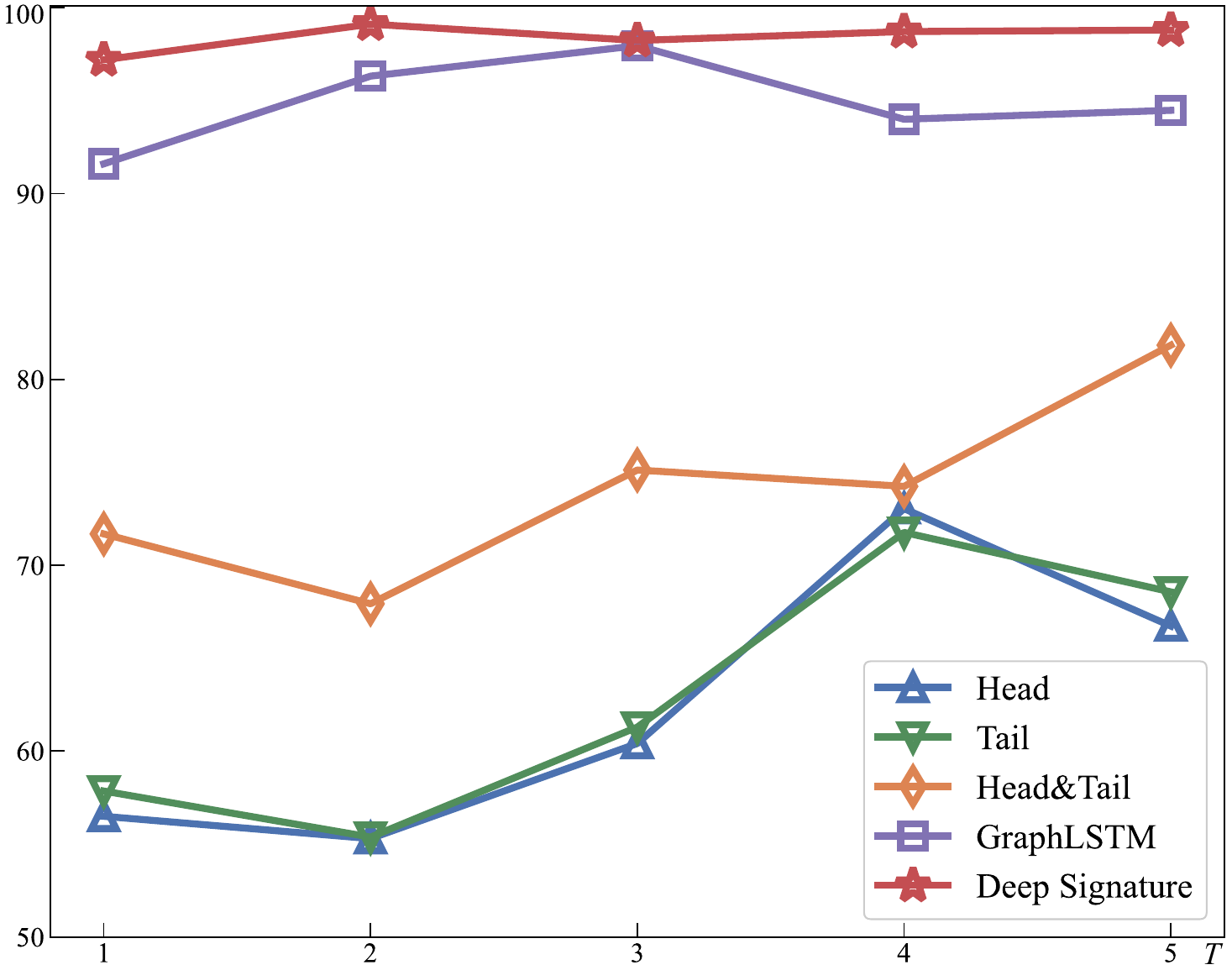}
    %\caption{fig1}
    }
    % \quad
    \subfigure[Different scales of system]{
    \includegraphics[width=.47\linewidth]{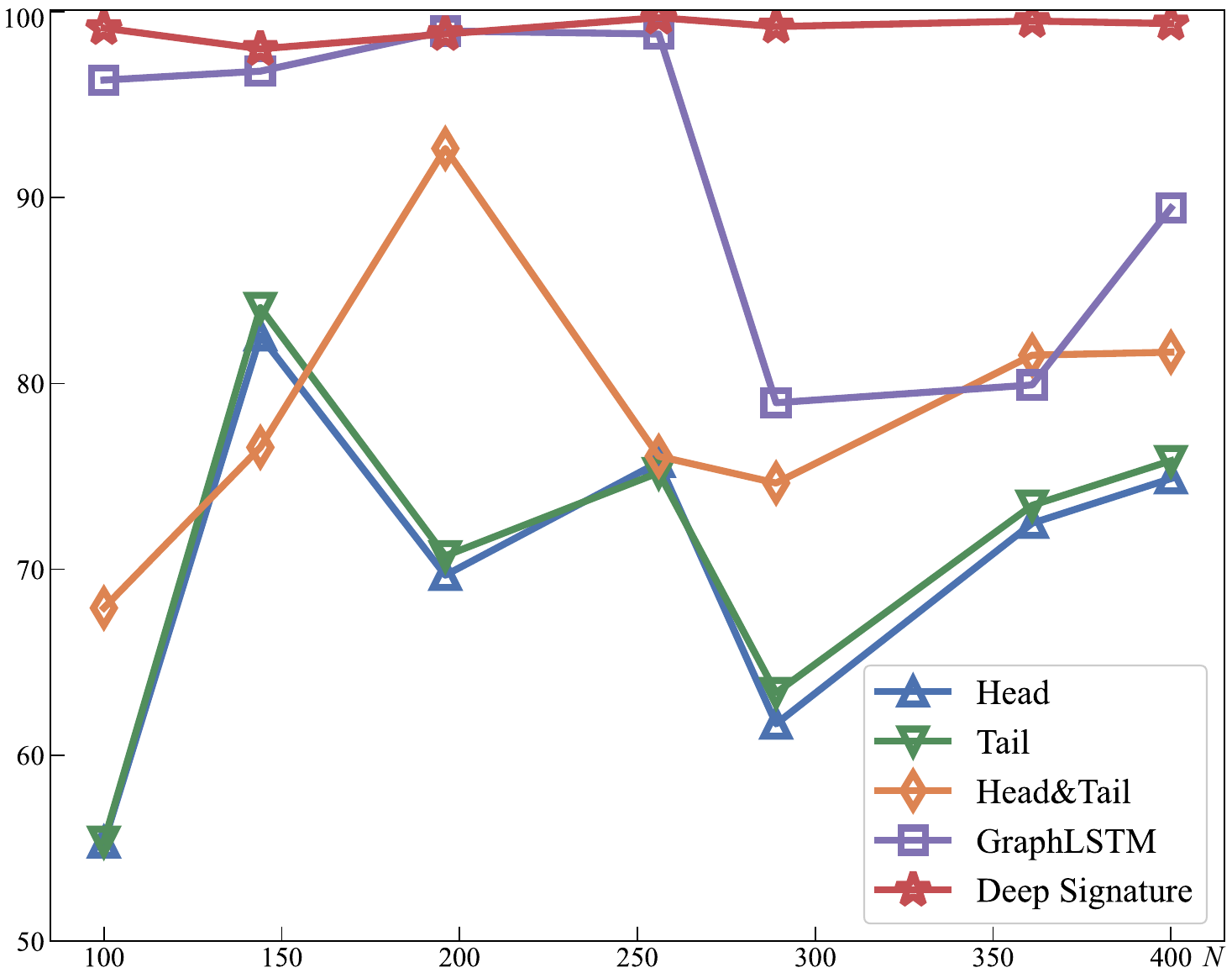}
    }
    \vspace{-0.4cm}
    \caption{The validation of scalability of our approach towards systems with various timescales (in a) and different number of atoms (in b).}
    \label{fig:exp_scale}
    \end{minipage}
\vspace{-0.2cm}
\end{figure}

For gene regulatory dynamics, we generate 100 trajectories that describe the interactive dynamics between genes and transcription factors. These dynamics are categorized into degradation type or dimerization type, with an equal number of trajectories for each type. The systems are simulated over a period of $2$s and with a time interval of $0.004$s, resulting in 500 frames per trajectory. Besides, each system encompasses 100 nodes with randomly generated Power-law network structure to describe their relationships. We implement the deep spectral clustering module in our approach using two graph pooling layers that first coarsen the dynamics into 60 nodes and then into 30 nodes. The coefficients of loss terms are set as $\lambda_1=1$, $\lambda_2=0.01$, and $\lambda_3=10$. We optimize our model using Adam with an initial learning rate of 5e-4 and a weight decay of 1e-4. For comparison, we consider several baseline approaches which aggregate nodes' dynamics by averaging without taking into account their interactions. These methods incorporate a deep spectral clustering model with an MLP to process the first frame only (Head), the last frame only (Tail), the first and last frame (Head \& Tail), \textcolor{black}{and all frames with an additional LSTM layer (GraphLSTM) or Transformer (GraphTrans) to aggregate nodes' temporal dynamics.} The overall comparison results with baseline methods are presented in Table~\ref{tab:exp_gene}.

From the table, we observe that our Deep Signature method achieves the best performance with a clear margin over baseline approaches. Besides, compared to GraphLSTM which aggregates atomic dynamics by averaging them directly, our Deep Signature exhibits a substantial performance boost by incorporating the log-signature transform into our method design, demonstrating the importance of capturing interatomic interaction dynamics for property prediction. Moreover, we validate the scalability of our approach to dynamical system across various time scales and different system sizes. As shown in Fig.~\ref{fig:exp_scale}(a), when extending the time duration of dynamics from $1$s to $5$s, our Deep Signature consistently obtains the best performance across these settings. Furthermore, when we change the system size via increasing the number of nodes that would enhance the complexity of dynamics, the results in Fig.~\ref{fig:exp_scale}(b) show that our method can still perform well under these conditions. These results showcase the good scalability of our method for dealing with dynamical systems of different timescales and sizes. In contrast, while GraphLSTM yields competitive results for systems with fewer than 250 nodes, a notable performance degradation appears as the number of nodes keeps increasing. We conjecture that this is because the averaged dynamics become less representative of the global dynamics as the system scale grows.

% -----------------------------------------------------------------------
\subsection{Experiments on epidermal growth factor receptor mutation dynamics}
\begin{table*}[!tbp]
\vspace{-0.1cm}
\caption{Comparison of different methods for the classification results on the EGFR dynamics.}
\vspace{-0.4cm}
\label{tab:exp_egfr}
\begin{center}
\fontsize{8.3}{9.3}\selectfont
% \small{
\begin{tabular}{p{36pt}<{\raggedright}p{32pt}<{\raggedright}p{32pt}<{\raggedright}p{30pt}<{\centering}p{30pt}<{\centering}p{32pt}<{\raggedright}p{32pt}<{\raggedright}p{32pt}<{\raggedright}p{38pt}<{\raggedright}}
\toprule[0.7pt]
\rot[25]{\textbf{Method}} & \rot[25][3em]{Dihedral angle} & \rot[25][3em]{C$\bm{\alpha}$-dihedral angle} & \rot[25][3em]{Head} &  \rot[25][3em]{Tail} & \rot[25][3em]{Head \& Tail} & \rot[25][3em]{GraphLSTM} & \rot[25][3em]{GraphTrans} & \rot[25][3em]{Deep Signature} \\
\hline  
\rule{0pt}{2ex} \textbf{Accuracy}  &67.47\scalebox{0.65}{$\pm0.98$}  &63.73\scalebox{0.65}{$\pm5.54$}  &61.47\scalebox{0.65}{$\pm9.43$}   &59.73\scalebox{0.65}{$\pm11.11$} &59.47\scalebox{0.65}{$\pm10.85$} &64.93\scalebox{0.65}{$\pm3.64$}   &64.40\scalebox{0.65}{$\pm6.55$} &~~69.33\scalebox{0.65}{$\pm4.78$} \\
\rule{0pt}{2ex} \textbf{Recall}    &~2.40\scalebox{0.65}{$\pm2.94$}  &~1.20\scalebox{0.65}{$\pm2.40$}   &~2.40\scalebox{0.65}{$\pm4.80$}    &~2.80\scalebox{0.65}{$\pm5.60$}  &~6.00\scalebox{0.65}{$\pm7.38$}  &11.60\scalebox{0.65}{$\pm8.52$}   &20.80\scalebox{0.65}{$\pm10.47$} &~~21.27\scalebox{0.65}{$\pm8.26$} \\ 
\bottomrule[0.7pt]
\end{tabular}
% }
\end{center}
% \vspace{-0.2cm}
\end{table*}

\begin{figure}[tbp]%{\textwidth}
    \begin{minipage}{.38\linewidth}
    % \fontsize{9}{9.5}\selectfont
    \centering
    \fontsize{8.3}{9.3}\selectfont
    \vspace{-0.1cm}
    \captionof{table}{Ablation study on different loss items for Deep Signature. We report accuracy and recall for performance evaluation.}
    \vspace{-0.2cm}
    % \small{
    \begin{tabular}{ccc|p{30pt}<{\centering}p{30pt}<{\centering}}
    \toprule[0.7pt]
    \rule{0pt}{2ex} $\mathcal{L}_u$ & $\mathcal{L}_t$ & $\mathcal{L}_c$ & \textbf{Accuracy} & \textbf{Recall} \\
    \hline  
    \rule{0pt}{2ex} $\times$     & $\times$      & $\checkmark$ & 67.60\scalebox{0.65}{$\pm6.02$} & 19.60\scalebox{0.65}{$\pm6.38$}\\
    \rule{0pt}{2ex} $\times$     & $\checkmark$  & $\checkmark$ & 67.97\scalebox{0.65}{$\pm1.15$} & 17.60\scalebox{0.65}{$\pm4.45$} \\
    \rule{0pt}{2ex} $\checkmark$ & $\times$      & $\checkmark$ & 66.00\scalebox{0.65}{$\pm2.23$} & 23.20\scalebox{0.65}{$\pm4.99$}\\
    \rule{0pt}{2ex} $\checkmark$ & $\checkmark$  & $\checkmark$ & 69.33\scalebox{0.65}{$\pm4.78$} & 21.27\scalebox{0.65}{$\pm8.26$}\\
    \bottomrule[0.7pt]
    \end{tabular}
    % }
    \label{tab:ablation_study}
    \end{minipage}\qquad\hfill
    \begin{minipage}{.58\linewidth}
    \vspace{-0.2cm}
    % \centering
    \includegraphics[width=0.98\textwidth]{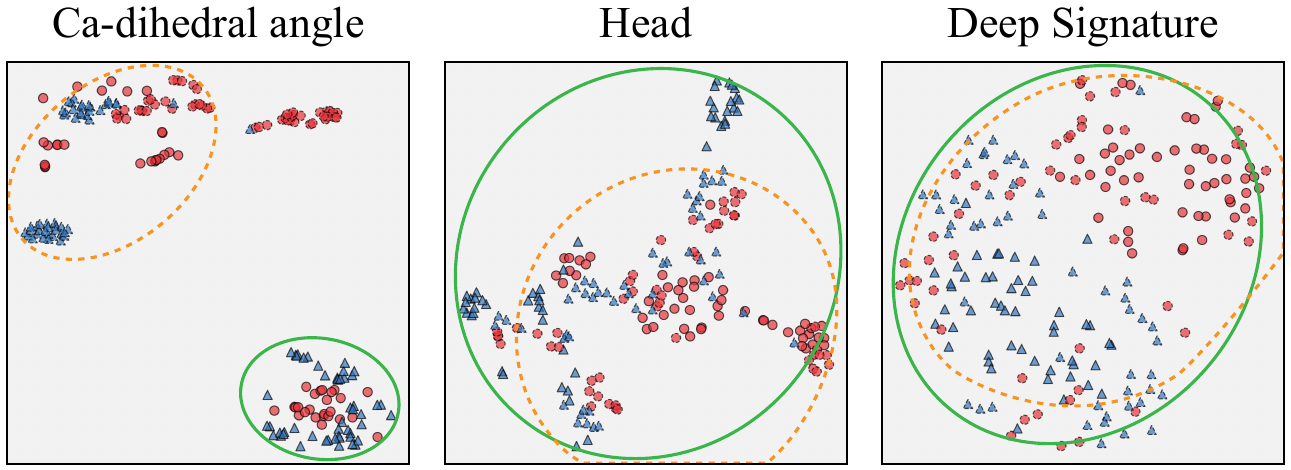}
    \vspace{-0.28cm}
    \caption{Visualization of the extracted features via t-SNE. Positive samples are presented in red circles and negative samples are in blue triangles.}
    \label{fig:emb_tsne}
    \end{minipage}
\vspace{-0.3cm}
\end{figure}
In this study, we investigate the binding process between EGFR mutations and EGFR-receptor tyrosine kinases (RTK) during which their interactions can contribute to drug resistance mechanisms. The combinations between four RTK partners and five mutation types together with wild type as a reference are considered. To acquire the dynamic data, we follow the pipeline presented in~\citep{Zhu2021BMC} that each system first undergoes energy minimization prior to simulation, then it is heated for 100 ps, followed by density equilibration for 100 ps and constant pressure equilibration for 5 ns, and finally runs simulation on the equilibrated structures for 50 ns. This results in 24 trajectories, each comprising 1,000 frames describing the temporal interactions among approximately 5,000 atoms. Each trajectory is labeled according to its sensitivity towards the drug. Here, our deep spectral clustering model consists of three layers that progressively coarsen the dynamics into 400, 200, 50 nodes. We set the scaling parameters as $\lambda_1=1$, $\lambda_2=0.01$, and $\lambda_3=10$. The model is trained for 200 epochs with an initial learning rate 5e-5 and a weight decay of 1e-5. In addition to the deep learning-based baselines introduced above, we further consider two baselines that extract dihedral angles as geometrical descriptors, followed by PCA to reduce the dimension of features and SVM for classification. The overall comparison results are summarized in Table~\ref{tab:exp_egfr}.
% Baseline: (1) geometrical characterization (2) free energy of binding (3) Deep Signature residue-level

As can be seen, our Deep Signature method consistently obtains the best classification results on EGFR dynamics, achieving an accuracy of $69.33\%$  and a recall of $21.27\%$. This validates the practical applicability of our method for its capability to tackle complex dynamics inherent in large-scale molecular systems. In addition, in contrast to baseline methods that achieve high accuracy but low recall which often due to their reliance on bypasses caused by the varying sample sizes for different classes, our method demonstrates a substantial improvement in recall, indicating its robustness under class imbalance issues. Furthermore, when compared to conventional geometrical descriptor methods, deep learning-based approaches, with the exception of ours, tend to underperform on this dataset. This discrepancy can be attributed to the intricate non-linearity and atomic fluctuations characteristic of EGFR dynamics, which pose significant challenges for deep learning techniques to extract meaningful patterns from a limited amount of data.

\noindent\textbf{Ablation study.}~We conduct an ablation study to assess the necessity of each loss term used in our model, with the results for EGFR dynamics presented in Table~\ref{tab:ablation_study}. We begin by training our Deep Signature model using only the classification loss $\mathcal{L}_c$, which yields an accuracy of $67.60\%$. Upon incorporating the temporal consistency loss $\mathcal{L}_t$, the accuracy increases to $67.97\%$. However, when we replace $\mathcal{L}_t$ with the spectral clustering loss $\mathcal{L}_u$, performance slightly degrades to $66.00\%$. This decrease may be attributed to the tendency of $\mathcal{L}_u$ to guide the spectral clustering module towards hard assignments, thereby impairing the representation ability of the coarse grained dynamics. Training with all three loss terms produces the best results, demonstrating the validity of our method design.

\begin{figure*}[!tbp]
\centering
\vspace{-0.2cm}
\includegraphics[width=\textwidth]{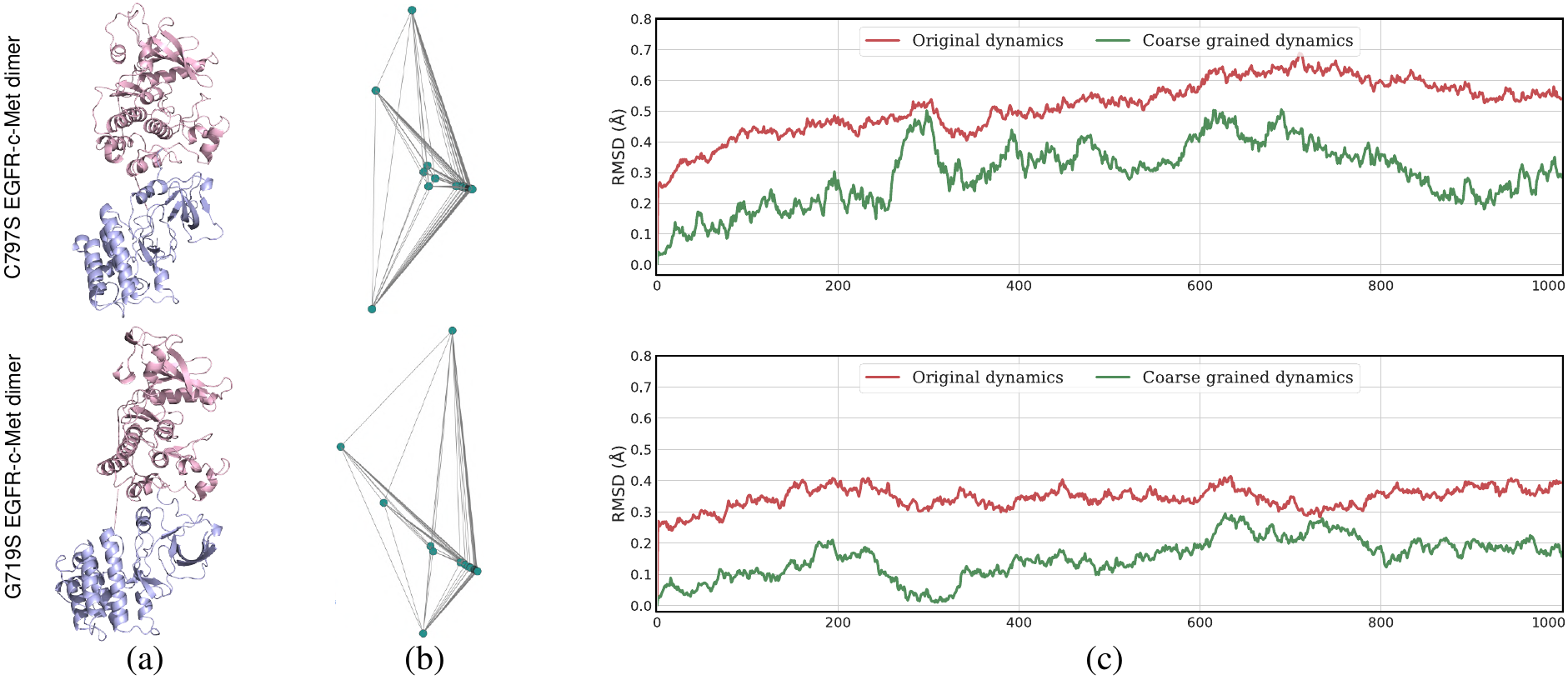}
\vspace{-0.65cm}
\caption{The effect of the deep spectral clustering module for C797S and G719S mutated EGFR dimers. (a) are origianl 3D conformations, (b) are coarse grained graphs, (c) are the RMSD curves of raw trajectories and coarse grained trajectories.}
\label{fig:cg_dynamics}
\vspace{-0.4cm}
\end{figure*}
\noindent\textbf{Analysis on deep spectral clustering module.}~To understand the impact of our spectral clustering module on the graph structures and trajectories for EGFR dynamics, we investigate two cases: a drug-resistant C797S mutant and a drug-sensitive G719S mutant respectively dimerized with EGFR partners, with their results visualized in Fig.~\ref{fig:cg_dynamics}. By comparing (a) and (b), we observe that coarse grained graphs produced by our module can maintain the overall structure of original conformations, and the nodes mainly distribute in the middle region, which is valid since interatomic interactions that facilitate protein function commonly occur in this area. Besides, as shown in (c), the root mean square deviation (RMSD) curves for both the raw and coarse grained trajectories generally follow the same trend, indicating a high fidelity of the coarse grained dynamics to the original dynamics.

% \noindent\textbf{The discriminability of features.}
\noindent\textbf{Analysis on path signature transform module.}~We visualize the feature space via t-SNE to explore the discriminability of learned features after path signature transform. The results of all data samples with colors indicating their classes are presented in Fig.~\ref{fig:emb_tsne}. As seen, both the non-learnable features, which compute dihedral angles for all C$\alpha$ atoms, and the learned features derived solely from the first frame tend to gather tightly in feature space, making it difficult to establish a decision boundary to distinguish samples belonging to different classes. In contrast, features extracted via our Deep Signature method are more uniformly distributed, and we can find such a decision boundary easily, verifying the good discriminability of these features.
% \noindent\textbf{Analysis on the heterogeneous distribution discrepancy.}~
We further use green circles to indicate training samples and orange dashed circles to denote test samples in Fig.~\ref{fig:emb_tsne}. These two circles typically exhibit limited overlap due to distribution shift between the training and test samples, which arises from heterogeneity in the number of atoms and molecular topological structures. Nevertheless, Deep Signature exhibits an ability to learn generalizable features for unseen samples. This is because learning features that respect symmetry enhances generalization capability~\citep{Schutt2017NC}, and incorporating layer normalization helps mitigate the distribution discrepancy~\citep{Ba2016LN}.
% This shift, which arises from heterogeneity in the number of atoms and molecular topological structures, poses great challenges for robust feature extraction.
% in our implementation 

\begin{figure*}[!tbp]
\centering
\includegraphics[width=\textwidth]{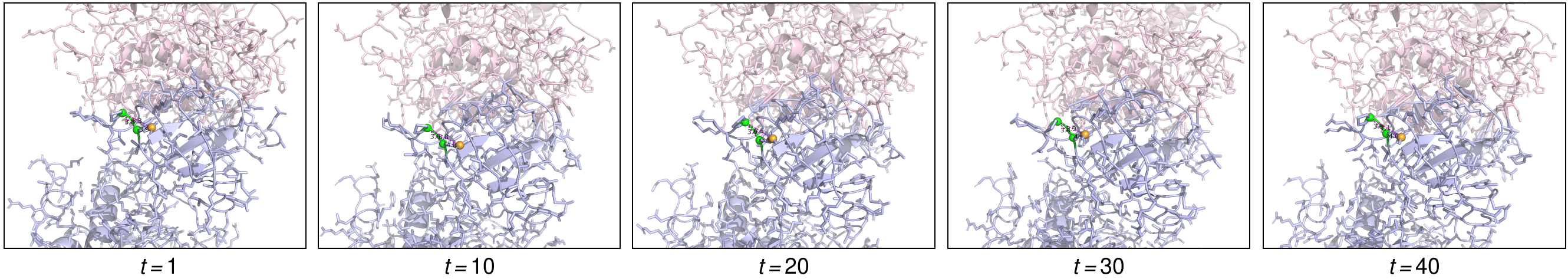}
\vspace{-0.7cm}
\caption{The visualization of critical pathways and interatomic interactions that contribute to drug sensitiveness on the EGFR dynamics.}
\label{fig:viz}
\vspace{-0.2cm}
\end{figure*}
\noindent\textbf{Analysis on interpretability.}~Our approach also exhibits notable interpretability for its ability to identify every possible type of interaction among atoms that are essential for protein functioning. This capability stems from the deep spectral clustering module which only involves linear mapping for cluster assignment, and log-signature features that maintain permutation equivariance as discussed in Section~\ref{sec:equivariance_invariance}. In practice, we apply the Gradient $\odot$ Input method~\citep{Shrikumar2017ICML} to quantify the contribution of each element in log-signature feature to the final prediction output and then identify three key atoms whose interactive dynamics play a pivotal role in the progression of EGFR. The identified atoms, along with their interatomic distances, are illustrated in Fig.~\ref{fig:viz}. Notably, these identified atoms fall within the hinge region that comprises all ATP binding sites, demonstrating strong consistency with the experimental observation~\citep{Kaufman2021Molecular}.
% Notably, these identified atoms fall within the tyrosine kinase domain of EGFR that contains all ATP binding sites, demonstrating strong consistency with the experimental observation~\citep{Kaufman2021Molecular}.

% -----------------------------------------------------------------------
\subsection{Experiments on G protein-coupled receptors dynamics}
% \begin{table*}[!tbp]
% \caption{Comparison of different methods for the classification results on the GPCR dynamics. }
% \vspace{-0.4cm}
% \label{tab:exp_gpcr}
% \begin{center}
% % \fontsize{8}{9.3}\selectfont
% \small{
%     \begin{tabular}{cp{36pt}<{\raggedright}p{36pt}<{\raggedright}p{34pt}<{\centering}p{34pt}<{\centering}p{36pt}<{\raggedright}p{36pt}<{\raggedright}p{44pt}<{\raggedright}}
%     \toprule[0.7pt]
%     \rot[25]{\textbf{Method}} & \rot[25][3em]{Dihedral angle} & \rot[25][3em]{C$\bm{\alpha}$-dihedral angle} & \rot[25][3em]{Head} &  \rot[25][3em]{Tail} & \rot[25][3em]{Head \& Tail} & \rot[25][3em]{GraphLSTM} & \rot[25][3em]{Deep Signature} \\
%     \hline
%     \rule{0pt}{2ex} \textbf{Accuracy} &50.00\scalebox{0.65}{$\pm0.00$} &44.00\scalebox{0.65}{$\pm1.12$}   &48.00\scalebox{0.65}{$\pm17.13$}  &46.67\scalebox{0.65}{$\pm16.33$} &48.53\scalebox{0.65}{$\pm18.83$} &53.33\scalebox{0.65}{$\pm12.47$}  &~~64.20\scalebox{0.65}{$\pm9.32$} \\
%     \rule{0pt}{2ex} \textbf{Recall}   &66.67\scalebox{0.65}{$\pm0.00$} &35.47\scalebox{0.65}{$\pm3.73$}  &29.33\scalebox{0.65}{$\pm7.35$}  &26.67\scalebox{0.65}{$\pm13.33$} &33.33\scalebox{0.65}{$\pm21.08$}  &26.66\scalebox{0.65}{$\pm13.33$}  &~~41.30\scalebox{0.65}{$\pm20.43$} \\ 
%     \bottomrule[0.7pt]
%     \end{tabular}
% }
% \end{center}
% \vspace{-0.2cm}
% \end{table*}
\begin{table*}[!tbp]
\caption{Comparison of different methods for the classification results on the GPCR dynamics. }
\vspace{-0.4cm}
\label{tab:exp_gpcr}
\begin{center}
\fontsize{8.3}{9.3}\selectfont
% \small{
    \begin{tabular}{p{36pt}<{\raggedright}p{32pt}<{\raggedright}p{32pt}<{\raggedright}p{30pt}<{\centering}p{30pt}<{\centering}p{32pt}<{\raggedright}p{32pt}<{\raggedright}p{32pt}<{\raggedright}p{38pt}<{\raggedright}}
    \toprule[0.7pt]
    \rot[25]{\textbf{Method}} & \rot[25][3em]{Dihedral angle} & \rot[25][3em]{C$\bm{\alpha}$-dihedral angle} & \rot[25][3em]{Head} &  \rot[25][3em]{Tail} & \rot[25][3em]{Head \& Tail} & \rot[25][3em]{GraphLSTM} & \rot[25][3em]{GraphTrans} & \rot[25][3em]{Deep Signature}  \\
    \hline
    \rule{0pt}{2ex} \textbf{Accuracy} &52.67\scalebox{0.65}{$\pm0.94$} &44.00\scalebox{0.65}{$\pm1.12$}   &53.07\scalebox{0.65}{$\pm6.47$}  &47.20\scalebox{0.65}{$\pm5.86$} &47.20\scalebox{0.65}{$\pm5.86$} &53.47\scalebox{0.65}{$\pm14.86$}  &44.80\scalebox{0.65}{$\pm6.97$} &~~58.00\scalebox{0.65}{$\pm4.17$} \\
    \rule{0pt}{2ex} \textbf{Recall}   &31.73\scalebox{0.65}{$\pm1.77$} &35.47\scalebox{0.65}{$\pm3.73$}  &26.40\scalebox{0.65}{$\pm13.21$}  &37.60\scalebox{0.65}{$\pm15.72$} &37.60\scalebox{0.65}{$\pm15.72$}  &38.13\scalebox{0.65}{$\pm13.19$}  &33.33\scalebox{0.65}{$\pm21.08$} &~~43.30\scalebox{0.65}{$\pm7.85$} \\ 
    \bottomrule[0.7pt]
    \end{tabular}
% }
\end{center}
\vspace{-0.2cm}
\end{table*}
The GPCR superfamily is a major therapeutic target as its functioning regulates nearly every physiological process in the human body. To create a dataset for its analysis, we select 26 structures with their MD simulations from the Molecular Dynamics Database for GPCRs (GPCRmd)~\citep{Rodriguez2020GPCR}, including 13 active state structures and 13 inactive state structures. The task is to identify these distinct active states. Each simulation is conducted for 500 ns with a time interval of 200 ps, therefore each trajectory consists of 2,500 frames that describe the conformational dynamics. Here, we only consider atoms in backbones. Experimental results are provided in Table~\ref{tab:exp_gpcr}.

It is evident that Deep Signature achieves the highest performance on the GPCR dynamics, with an accuracy of $58.00\%$ and a recall ratio of $43.30\%$. The results are significantly better than the compared baselines. Traditional geometrical descriptor methods, such as the dihedral angles and C$\alpha$-dihedral angles, exhibit relatively lower classification accuracy and recall compared to our method. This again verifies that geometrical descriptors are inadequate for capturing dynamic patterns when the dynamics are highly nonlinear due to the existence of atomic fluctuations in GPCR dynamics.
Moreover, relying solely on the first or last frame fails to account for the dynamic nature of interactions, thereby limiting the ability to discern useful patterns for predicting structural states. Although the GraphLSTM method outperforms the geometric descriptor methods, achieving a moderate accuracy of $53.47\%$ and a recall ratio of $38.13\%$, it still falls short of the superior classification capabilities exhibited by the Deep Signature method, demonstrating the effectiveness of our approach in characterizing the complex dynamics of GPCR systems.

% =======================================================================
\section{Conclusion}
\label{sec:conclusion}
In this paper, we introduce a novel deep learning framework, Deep Signature, to deal with the long-standing challenge of understanding protein dynamics in large-scale biological systems. It comprises soft spectral clustering to aggregate cooperative dynamics and log-signature transformation to characterize global interactive dynamics. Theoretically, our method exhibits desirable properties such as invariance to translation, near-invariance to rotation, and equivariance to atomic coordinate permutation. Experimental results on three biological process benchmarks verify the effectiveness of Deep Signature in capturing the complex interactive dynamics of large-scale molecular systems. We hope our work can offer a promising new direction for the analysis of protein dynamics.


\section*{Acknowledgments}
This work was  supported  in part by the Hong Kong Innovation and Technology Commission (ITC) (InnoHK Project CIMDA),  in part by the Institute of Digital Medicine of City University of Hong Kong (Project 9229503), in part by the Hong Kong Research Grants Council under Projects 21200522, 11200323 and 11203220, in part by
Chow Sang Sang Donation and Matching Fund (Project 9229161), and in part by the Hong Kong Innovation and Technology Commission (Project GHP/044/21SZ).

% \newpage
\bibliography{egbib}

\begin{thebibliography}{48}
\providecommand{\natexlab}[1]{#1}
\providecommand{\url}[1]{\texttt{#1}}
\expandafter\ifx\csname urlstyle\endcsname\relax
  \providecommand{\doi}[1]{doi: #1}\else
  \providecommand{\doi}{doi: \begingroup \urlstyle{rm}\Url}\fi

\bibitem[Araki et~al.(2021)Araki, Matsumoto, Bekker, Isaka, Sagae, Kamiya, and Okuno]{Araki2021NC}
Mitsugu Araki, Shigeyuki Matsumoto, Gert-Jan Bekker, Yuta Isaka, Yukari Sagae, Narutoshi Kamiya, and Yasushi Okuno.
\newblock Exploring ligand binding pathways on proteins using hypersound-accelerated molecular dynamics.
\newblock \emph{Nature communications}, 12\penalty0 (1):\penalty0 2793, 2021.

\bibitem[Ba et~al.(2016)Ba, Kiros, and Hinton]{Ba2016LN}
Jimmy~Lei Ba, Jamie~Ryan Kiros, and Geoffrey~E. Hinton.
\newblock Layer normalization, 2016.
\newblock URL \url{https://arxiv.org/abs/1607.06450}.

\bibitem[Batatia et~al.(2022)Batatia, Kovacs, Simm, Ortner, and Cs{\'a}nyi]{Batatia2022NIPS}
Ilyes Batatia, David~P Kovacs, Gregor Simm, Christoph Ortner, and G{\'a}bor Cs{\'a}nyi.
\newblock Mace: Higher order equivariant message passing neural networks for fast and accurate force fields.
\newblock \emph{Advances in Neural Information Processing Systems}, 35:\penalty0 11423--11436, 2022.

\bibitem[Batzner et~al.(2022)Batzner, Musaelian, Sun, Geiger, Mailoa, Kornbluth, Molinari, Smidt, and Kozinsky]{Batzner2022NC}
Simon Batzner, Albert Musaelian, Lixin Sun, Mario Geiger, Jonathan~P Mailoa, Mordechai Kornbluth, Nicola Molinari, Tess~E Smidt, and Boris Kozinsky.
\newblock E (3)-equivariant graph neural networks for data-efficient and accurate interatomic potentials.
\newblock \emph{Nature communications}, 13\penalty0 (1):\penalty0 2453, 2022.

\bibitem[Bianchi et~al.(2020)Bianchi, Grattarola, and Alippi]{Bianchi2020ICML}
Filippo~Maria Bianchi, Daniele Grattarola, and Cesare Alippi.
\newblock Spectral clustering with graph neural networks for graph pooling.
\newblock In \emph{International conference on machine learning}, pp.\  874--883. PMLR, 2020.

\bibitem[Car \& Parrinello(1985)Car and Parrinello]{Car1985PRL}
Richard Car and Mark Parrinello.
\newblock Unified approach for molecular dynamics and density-functional theory.
\newblock \emph{Physical review letters}, 55\penalty0 (22):\penalty0 2471, 1985.

\bibitem[Choi \& Park(2023)Choi and Park]{Choi2023TIST}
Jeongwhan Choi and Noseong Park.
\newblock Graph neural rough differential equations for traffic forecasting.
\newblock \emph{ACM Transactions on Intelligent Systems and Technology}, 14\penalty0 (4):\penalty0 1--27, 2023.

\bibitem[Diehl(2013)]{Diehl2013arXiv}
Joscha Diehl.
\newblock Rotation invariants of two dimensional curves based on iterated integrals.
\newblock \emph{arXiv preprint arXiv:1305.6883}, 2013.

\bibitem[Dror et~al.(2012)Dror, Dirks, Grossman, Xu, and Shaw]{Dror2012ARB}
Ron~O Dror, Robert~M Dirks, JP~Grossman, Huafeng Xu, and David~E Shaw.
\newblock Biomolecular simulation: a computational microscope for molecular biology.
\newblock \emph{Annual review of biophysics}, 41:\penalty0 429--452, 2012.

\bibitem[Endo et~al.(2019)Endo, Yuhara, Tomobe, and Yasuoka]{Endo2019Nano}
Katsuhiro Endo, Daisuke Yuhara, Katsufumi Tomobe, and Kenji Yasuoka.
\newblock Detection of molecular behavior that characterizes systems using a deep learning approach.
\newblock \emph{Nanoscale}, 11\penalty0 (20):\penalty0 10064--10071, 2019.

\bibitem[Frenkel \& Smit(2023)Frenkel and Smit]{Frenkel2023Understanding}
Daan Frenkel and Berend Smit.
\newblock \emph{Understanding molecular simulation: from algorithms to applications}.
\newblock Elsevier, 2023.

\bibitem[Gainza et~al.(2020{\natexlab{a}})Gainza, Sverrisson, Monti, Rodola, Boscaini, Bronstein, and Correia]{Gainza2020MaSIF}
Pablo Gainza, Freyr Sverrisson, Frederico Monti, Emanuele Rodola, Davide Boscaini, Michael~M Bronstein, and Bruno~E Correia.
\newblock Deciphering interaction fingerprints from protein molecular surfaces using geometric deep learning.
\newblock \emph{Nature Methods}, 17\penalty0 (2):\penalty0 184--192, 2020{\natexlab{a}}.

\bibitem[Gainza et~al.(2020{\natexlab{b}})Gainza, Sverrisson, Monti, Rodola, Boscaini, Bronstein, and Correia]{Gainza2020NM}
Pablo Gainza, Freyr Sverrisson, Frederico Monti, Emanuele Rodola, Davide Boscaini, Michael~M Bronstein, and Bruno~E Correia.
\newblock Deciphering interaction fingerprints from protein molecular surfaces using geometric deep learning.
\newblock \emph{Nature Methods}, 17\penalty0 (2):\penalty0 184--192, 2020{\natexlab{b}}.

\bibitem[Gao et~al.(2016)Gao, Barzel, and Barab{\'a}si]{Gao2016Nature}
Jianxi Gao, Baruch Barzel, and Albert-L{\'a}szl{\'o} Barab{\'a}si.
\newblock Universal resilience patterns in complex networks.
\newblock \emph{Nature}, 530\penalty0 (7590):\penalty0 307--312, 2016.

\bibitem[Gao et~al.(2020)Gao, Nguyen, Sresht, Mathiowetz, Tu, and Wei]{Gao2020PCCP}
Kaifu Gao, Duc~Duy Nguyen, Vishnu Sresht, Alan~M Mathiowetz, Meihua Tu, and Guo-Wei Wei.
\newblock Are 2d fingerprints still valuable for drug discovery?
\newblock \emph{Physical chemistry chemical physics}, 22\penalty0 (16):\penalty0 8373--8390, 2020.

\bibitem[Ing{\'o}lfsson et~al.(2014)Ing{\'o}lfsson, Lopez, Uusitalo, de~Jong, Gopal, Periole, and Marrink]{Ingolfsson2014CMS}
Helgi~I Ing{\'o}lfsson, Cesar~A Lopez, Jaakko~J Uusitalo, Djurre~H de~Jong, Srinivasa~M Gopal, Xavier Periole, and Siewert~J Marrink.
\newblock The power of coarse graining in biomolecular simulations.
\newblock \emph{Wiley Interdisciplinary Reviews: Computational Molecular Science}, 4\penalty0 (3):\penalty0 225--248, 2014.

\bibitem[Jin et~al.(2022)Jin, Pak, Durumeric, Loose, and Voth]{Jin2022JCTC}
Jaehyeok Jin, Alexander~J Pak, Aleksander~EP Durumeric, Timothy~D Loose, and Gregory~A Voth.
\newblock Bottom-up coarse-graining: Principles and perspectives.
\newblock \emph{Journal of Chemical Theory and Computation}, 18\penalty0 (10):\penalty0 5759--5791, 2022.

\bibitem[Kaufman et~al.(2021)Kaufman, Dhingra, Jois, and Vicente]{Kaufman2021Molecular}
Nichole~EM Kaufman, Simran Dhingra, Seetharama~D Jois, and Maria da Graca~H Vicente.
\newblock Molecular targeting of epidermal growth factor receptor (egfr) and vascular endothelial growth factor receptor (vegfr).
\newblock \emph{Molecules}, 26\penalty0 (4):\penalty0 1076, 2021.

\bibitem[Khot et~al.(2019)Khot, Shiring, and Savoie]{Khot2019JCP}
Aditi Khot, Stephen~B Shiring, and Brett~M Savoie.
\newblock Evidence of information limitations in coarse-grained models.
\newblock \emph{The Journal of chemical physics}, 151\penalty0 (24), 2019.

\bibitem[Kidger \& Lyons(2021)Kidger and Lyons]{Kidger2021ICLR}
Patrick Kidger and Terry Lyons.
\newblock Signatory: differentiable computations of the signature and logsignature transforms, on both {\{}cpu{\}} and {\{}gpu{\}}.
\newblock In \emph{International Conference on Learning Representations}, 2021.
\newblock URL \url{https://openreview.net/forum?id=lqU2cs3Zca}.

\bibitem[Law et~al.(2017)Law, Sapienza, Zhang, Zuo, and Petit]{Law2017JACS}
Anthony~B Law, Paul~J Sapienza, Jun Zhang, Xiaobing Zuo, and Chad~M Petit.
\newblock Native state volume fluctuations in proteins as a mechanism for dynamic allostery.
\newblock \emph{Journal of the American Chemical Society}, 139\penalty0 (10):\penalty0 3599--3602, 2017.

\bibitem[Lewandowski et~al.(2015)Lewandowski, Halse, Blackledge, and Emsley]{Lewandowski2015Science}
J{\'o}zef~R Lewandowski, Meghan~E Halse, Martin Blackledge, and Lyndon Emsley.
\newblock Direct observation of hierarchical protein dynamics.
\newblock \emph{Science}, 348\penalty0 (6234):\penalty0 578--581, 2015.

\bibitem[Li et~al.(2020{\natexlab{a}})Li, Yang, Capra, and Gerstein]{Li2020PLoS}
Bian Li, Yucheng~T Yang, John~A Capra, and Mark~B Gerstein.
\newblock Predicting changes in protein thermodynamic stability upon point mutation with deep 3d convolutional neural networks.
\newblock \emph{PLoS computational biology}, 16\penalty0 (11):\penalty0 e1008291, 2020{\natexlab{a}}.

\bibitem[Li et~al.(2022)Li, Liu, Chen, Yuan, Yu, Gou, Guo, and Pu]{Li2022JCIM}
Chuan Li, Jiangting Liu, Jianfang Chen, Yuan Yuan, Jin Yu, Qiaolin Gou, Yanzhi Guo, and Xuemei Pu.
\newblock An interpretable convolutional neural network framework for analyzing molecular dynamics trajectories: A case study on functional states for g-protein-coupled receptors.
\newblock \emph{Journal of Chemical Information and Modeling}, 62\penalty0 (6):\penalty0 1399--1410, 2022.

\bibitem[Li et~al.(2020{\natexlab{b}})Li, Wellawatte, Chakraborty, Gandhi, Xu, and White]{Li2020CS}
Zhiheng Li, Geemi~P Wellawatte, Maghesree Chakraborty, Heta~A Gandhi, Chenliang Xu, and Andrew~D White.
\newblock Graph neural network based coarse-grained mapping prediction.
\newblock \emph{Chemical science}, 11\penalty0 (35):\penalty0 9524--9531, 2020{\natexlab{b}}.

\bibitem[Li et~al.(2024)Li, Huang, Xia, Patterson, and Hong]{Li2024Bio}
Zoe Li, Ruili Huang, Menghang Xia, Tucker~A Patterson, and Huixiao Hong.
\newblock Fingerprinting interactions between proteins and ligands for facilitating machine learning in drug discovery.
\newblock \emph{Biomolecules}, 14\penalty0 (1):\penalty0 72, 2024.

\bibitem[Lyons(2014)]{Terry2014ICM}
Terry Lyons.
\newblock Rough paths, signatures and the modelling of functions on streams.
\newblock \emph{International Congress of Mathematicians-Seoul}, IV, 2014.

\bibitem[Lyons et~al.(2007)Lyons, Caruana, and L{\'e}vy]{Terry2007Springer}
Terry~J Lyons, Michael Caruana, and Thierry L{\'e}vy.
\newblock \emph{Differential equations driven by rough paths}.
\newblock Springer, 2007.

\bibitem[Majewski et~al.(2023)Majewski, P{\'e}rez, Th{\"o}lke, Doerr, Charron, Giorgino, Husic, Clementi, No{\'e}, and De~Fabritiis]{Majewski2023NC}
Maciej Majewski, Adri{\`a} P{\'e}rez, Philipp Th{\"o}lke, Stefan Doerr, Nicholas~E Charron, Toni Giorgino, Brooke~E Husic, Cecilia Clementi, Frank No{\'e}, and Gianni De~Fabritiis.
\newblock Machine learning coarse-grained potentials of protein thermodynamics.
\newblock \emph{Nature Communications}, 14\penalty0 (1):\penalty0 5739, 2023.

\bibitem[Marrink \& Tieleman(2013)Marrink and Tieleman]{Marrink2013CSR}
Siewert~J Marrink and D~Peter Tieleman.
\newblock Perspective on the martini model.
\newblock \emph{Chemical Society Reviews}, 42\penalty0 (16):\penalty0 6801--6822, 2013.

\bibitem[Mustali et~al.(2023)Mustali, Yasuda, Hirano, Yasuoka, Gautieri, and Arai]{Mustali2023RSC}
Jessica Mustali, Ikki Yasuda, Yoshinori Hirano, Kenji Yasuoka, Alfonso Gautieri, and Noriyoshi Arai.
\newblock Unsupervised deep learning for molecular dynamics simulations: a novel analysis of protein--ligand interactions in sars-cov-2 m pro.
\newblock \emph{RSC advances}, 13\penalty0 (48):\penalty0 34249--34261, 2023.

\bibitem[Otten et~al.(2018)Otten, Liu, Kenner, Clarkson, Mavor, Tawfik, Kern, and Fraser]{Otten2018NC}
Renee Otten, Lin Liu, Lillian~R Kenner, Michael~W Clarkson, David Mavor, Dan~S Tawfik, Dorothee Kern, and James~S Fraser.
\newblock Rescue of conformational dynamics in enzyme catalysis by directed evolution.
\newblock \emph{Nature communications}, 9\penalty0 (1):\penalty0 1314, 2018.

\bibitem[Qiu et~al.(2023)Qiu, O’Connor, Xue, Liu, and Huang]{Qiu2023JCTC}
Yunrui Qiu, Michael~S O’Connor, Mingyi Xue, Bojun Liu, and Xuhui Huang.
\newblock An efficient path classification algorithm based on variational autoencoder to identify metastable path channels for complex conformational changes.
\newblock \emph{Journal of Chemical Theory and Computation}, 19\penalty0 (14):\penalty0 4728--4742, 2023.

\bibitem[Reizenstein(2017)]{Reizenstein2017arXiv}
Jeremy Reizenstein.
\newblock Calculation of iterated-integral signatures and log signatures.
\newblock \emph{arXiv preprint arXiv:1712.02757}, 2017.

\bibitem[Rodr{\'\i}guez-Espigares et~al.(2020)Rodr{\'\i}guez-Espigares, Torrens-Fontanals, Tiemann, Aranda-Garc{\'\i}a, Ram{\'\i}rez-Anguita, Stepniewski, Worp, Varela-Rial, Morales-Pastor, Medel-Lacruz, et~al.]{Rodriguez2020GPCR}
Ismael Rodr{\'\i}guez-Espigares, Mariona Torrens-Fontanals, Johanna~KS Tiemann, David Aranda-Garc{\'\i}a, Juan~Manuel Ram{\'\i}rez-Anguita, Tomasz~Maciej Stepniewski, Nathalie Worp, Alejandro Varela-Rial, Adri{\'a}n Morales-Pastor, Brian Medel-Lacruz, et~al.
\newblock Gpcrmd uncovers the dynamics of the 3d-gpcrome.
\newblock \emph{Nature Methods}, 17\penalty0 (8):\penalty0 777--787, 2020.

\bibitem[Rogers et~al.(2023)Rogers, Nikol{\'e}nyi, and AlQuraishi]{Rogers2023Growing}
Julia~R Rogers, Gerg{\H{o}} Nikol{\'e}nyi, and Mohammed AlQuraishi.
\newblock Growing ecosystem of deep learning methods for modeling protein--protein interactions.
\newblock \emph{Protein Engineering, Design and Selection}, 36:\penalty0 gzad023, 2023.

\bibitem[Sch{\"u}tt et~al.(2017{\natexlab{a}})Sch{\"u}tt, Kindermans, Sauceda~Felix, Chmiela, Tkatchenko, and M{\"u}ller]{Schutt2017NIPS}
Kristof Sch{\"u}tt, Pieter-Jan Kindermans, Huziel~Enoc Sauceda~Felix, Stefan Chmiela, Alexandre Tkatchenko, and Klaus-Robert M{\"u}ller.
\newblock Schnet: A continuous-filter convolutional neural network for modeling quantum interactions.
\newblock \emph{Advances in neural information processing systems}, 30, 2017{\natexlab{a}}.

\bibitem[Sch{\"u}tt et~al.(2017{\natexlab{b}})Sch{\"u}tt, Arbabzadah, Chmiela, M{\"u}ller, and Tkatchenko]{Schutt2017NC}
Kristof~T Sch{\"u}tt, Farhad Arbabzadah, Stefan Chmiela, Klaus~R M{\"u}ller, and Alexandre Tkatchenko.
\newblock Quantum-chemical insights from deep tensor neural networks.
\newblock \emph{Nature communications}, 8\penalty0 (1):\penalty0 13890, 2017{\natexlab{b}}.

\bibitem[Shrikumar et~al.(2017)Shrikumar, Greenside, and Kundaje]{Shrikumar2017ICML}
Avanti Shrikumar, Peyton Greenside, and Anshul Kundaje.
\newblock Learning important features through propagating activation differences.
\newblock In \emph{International conference on machine learning}, pp.\  3145--3153. PMlR, 2017.

\bibitem[Sun et~al.(2023)Sun, Huang, Li, Gong, and Ye]{Sun2023DSR}
Daiwen Sun, He~Huang, Yao Li, Xinqi Gong, and Qiwei Ye.
\newblock Dsr: Dynamical surface representation as implicit neural networks for protein.
\newblock In \emph{Advances in Neural Information Processing Systems}, volume~36, pp.\  13873--13886, 2023.

\bibitem[Thaler et~al.(2022)Thaler, Stupp, and Zavadlav]{Thaler2022JCP}
Stephan Thaler, Maximilian Stupp, and Julija Zavadlav.
\newblock Deep coarse-grained potentials via relative entropy minimization.
\newblock \emph{The Journal of Chemical Physics}, 157\penalty0 (24), 2022.

\bibitem[Toprak et~al.(2012)Toprak, Veres, Michel, Chait, Hartl, and Kishony]{Toprak2012NG}
Erdal Toprak, Adrian Veres, Jean-Baptiste Michel, Remy Chait, Daniel~L Hartl, and Roy Kishony.
\newblock Evolutionary paths to antibiotic resistance under dynamically sustained drug selection.
\newblock \emph{Nature genetics}, 44\penalty0 (1):\penalty0 101--105, 2012.

\bibitem[Tsai et~al.(2009)Tsai, Del~Sol, and Nussinov]{Tsai2009MB}
Chung-Jung Tsai, Antonio Del~Sol, and Ruth Nussinov.
\newblock Protein allostery, signal transmission and dynamics: a classification scheme of allosteric mechanisms.
\newblock \emph{Molecular Biosystems}, 5\penalty0 (3):\penalty0 207--216, 2009.

\bibitem[Wang \& G{\'o}mez-Bombarelli(2019)Wang and G{\'o}mez-Bombarelli]{Wang2019NPJ}
Wujie Wang and Rafael G{\'o}mez-Bombarelli.
\newblock Coarse-graining auto-encoders for molecular dynamics.
\newblock \emph{npj Computational Materials}, 5\penalty0 (1):\penalty0 125, 2019.

\bibitem[Wang \& Zang(2024)Wang and Zang]{Wang2024MFG}
Zengqiang Wang and Di~Zang.
\newblock Multi-frequency graph neural rough differential equations for traffic forecasting.
\newblock 2024.

\bibitem[Webb et~al.(2018)Webb, Delannoy, and De~Pablo]{Webb2018JCTC}
Michael~A Webb, Jean-Yves Delannoy, and Juan~J De~Pablo.
\newblock Graph-based approach to systematic molecular coarse-graining.
\newblock \emph{Journal of chemical theory and computation}, 15\penalty0 (2):\penalty0 1199--1208, 2018.

\bibitem[Yasuda et~al.(2022)Yasuda, Endo, Yamamoto, Hirano, and Yasuoka]{Yasuda2022CB}
Ikki Yasuda, Katsuhiro Endo, Eiji Yamamoto, Yoshinori Hirano, and Kenji Yasuoka.
\newblock Differences in ligand-induced protein dynamics extracted from an unsupervised deep learning approach correlate with protein--ligand binding affinities.
\newblock \emph{Communications biology}, 5\penalty0 (1):\penalty0 481, 2022.

\bibitem[Zhu et~al.(2021)Zhu, Wang, and Yan]{Zhu2021BMC}
Mengxu Zhu, Debby~D Wang, and Hong Yan.
\newblock Genotype-determined egfr-rtk heterodimerization and its effects on drug resistance in lung cancer treatment revealed by molecular dynamics simulations.
\newblock \emph{BMC Molecular and Cell Biology}, 22\penalty0 (1):\penalty0 34, 2021.

\end{thebibliography}
\bibliographystyle{iclr2025_conference}

\newpage
\appendix

% \section{Appendix}

\noindent\textbf{{\large Table of Contents}}

\begin{itemize}
    \item[] \hyperref[app:property_proof]{\textbf{Appendix A:}} Proof for Properties
    
    \item[] \hyperref[app:exp_setup]{\textbf{Appendix B:}} Experimental Settings

    \item[] \hyperref[sec:app_exp_result]{\textbf{Appendix C:}} Further Results and Analysis

\end
{itemize}
\vspace{0.2cm}

% \section{Supplementary Material}

% Authors may wish to optionally include extra information (complete proofs, additional experiments and plots) in the appendix. All such materials should be part of the supplemental material (submitted separately) and should \textcolor{red}{NOT} be included in the main submission.

% =======================================================================
\section{Proof for Properties}
\label{app:property_proof}
In the beginning of our proof, we would first decompose our feature extracting process into two operations $g_{\mathrm{GNN}}(\cdot)$ and $\mathrm{LogSig}(\cdot)$ as presented in Section~\ref{sec:method} such that the overall feature transform is their composition $g_{\mathrm{GNN}}\circ \mathrm{LogSig}$. This decomposition offers us the opportunity to analyze the properties of component separately. Besides, since the logarithm map is bijective, which implies there is one-to-one correspondence between the signature and the log-signature~\citep{Terry2007Springer}, thus we can turn to analyze signature transform for intuitive demonstration.

% \textcolor{orange}{A function $f$ is said to be equivariant to the action of a group $G$ if $T_g(f(x))=f(S_g(x))$ for all $g \in G$, where $S_g$, $T_g$ are linear representations related to the group element $g$ (Serre, 1977).\\
% Serre, J.-P. Linear representations of finite groups, volume 42. Springer, 1977. \\
% (Equivariant Diffusion for Molecule Generation in 3D, Section 2.2)}

% -----------------------------------------------------------------------
\subsection{Translation Invariance}
Given trajectory data $\mathbf{X}_{1:T}\in\RR^{T\times N\times 3}$, let $\mathcal{T}_B$ represent a translation matrix $B\in\RR^{N\times 3}$ on the trajectory data such that at a certain time stamp $t$ we have $\mathcal{T}_{B}(\mathbf{X}_t)=\mathbf{X}_t+B$. For the coarse grained dynamics acquired by $\tilde{\mathbf{X}}_t^{\mathrm{pool}} = \mathbf{Q}^T\mathbf{X}_t$ as presented by Eq.~(\ref{eq:coarse_grain}), as $\mathbf{Q}^T$ is a linear matrix, we can get that $\mathbf{Q}^T \mathcal{T}_B(\mathbf{X}_t) = \mathbf{Q}^T\mathbf{X}_t+\mathbf{Q}^TB= \mathcal{T}_{\mathbf{Q}^TB}(\mathbf{Q}^T\mathbf{X}_t)$, thus $\tilde{\mathbf{X}}_{1:T}$ maintain equivariance with respect to translation $\mathcal{T}_B$ on input trajectory $\mathbf{X}_{1:T}$. Besides, as presented in Eq.~(\ref{eq:sig_integral}) that the path signature is composed of iterated path integrals, it inherits the properties of translation invariance that $\mathrm{Sig}^D_{a,b}(\tilde{\mathbf{X}}+c)=\mathrm{Sig}^D_{a,b}(\tilde{\mathbf{X}})$. In the whole, our path signature features $\mathrm{Sig}^D_{a,b}(\tilde{\mathbf{X}})$ are translation-invariant with respect to input trajectory data $\mathbf{X}_{a,b}$.

% -----------------------------------------------------------------------
\subsection{Rotation Invariance}
Given the coarse grained dynamics acquired by $\tilde{\mathbf{X}}_t = \mathbf{Q}^T\mathbf{X}_t$, where $\mathbf{X}_t\in\RR^{N\times 3}$ and $\tilde{\mathbf{X}}_t\in\RR^{M\times 3}$. The cluster assignment matrix $\mathbf{Q}\in\RR^{N\times M}$ holds that $\mathbf{Q}\mathbf{1}_M=\mathbf{1}_N$ and $\mathbf{Q}^T\mathbf{Q}=\mathbf{I}_M$. For any orthogonal matrix $\mathbf{R}_\theta\in\RR^{3\times 3}$ performed on $\mathbf{X}_t$ termed as $\mathbf{R}_\theta\mathbf{X}_t=(\mathbf{R}_\theta\mathbf{X}_t^1,\dots,\mathbf{R}_\theta\mathbf{X}_t^N)$, we proof below that rotating the input results in an equivalent rotation of the output $\mathbf{R}_\phi\tilde{\mathbf{X}}_t=\mathbf{Q}^T\mathbf{R}_\theta\mathbf{X}_t$, where $\mathbf{R}_{\phi}\in\RR^{M\times 3}$ is an orthogonal matrix performed on $\tilde{\mathbf{X}}_t$. Let we start with the right-hand side $\mathbf{Q}^T\mathbf{R}_\theta\mathbf{X}_t$. Here, $\mathbf{Q}^T$ has the form $\mathbf{Q}^T=(\mathbf{Q}^T_1,\dots,\mathbf{Q}^T_M)$, where the assignment function for $m$-th cluster is $\mathbf{Q}^T_m\in\RR^{N}$, $m\in\{1,\dots,M\}$, and the corresponding aggregated coordinate is $\mathbf{Q}^T_m\mathbf{X}_t$ which is a linear combination in each dimension. Due to the fact that $\mathbf{Q}^T_m\mathbf{R}_\theta\mathbf{X}_t$ can always be represented by $\mathbf{R}_{\phi}\mathbf{Q}^T_m\mathbf{X}_t$, which implies that for a rotation transform $\mathbf{R}_\theta$ on the original conformational space, there always exists an equivalent rotation transformation $\mathbf{R}_{\phi}$ in coarse grained conformational space. We then accomplish our proof that $\mathbf{R}_\phi\tilde{\mathbf{X}}_t=\mathbf{Q}^T\mathbf{R}_\theta\mathbf{X}_t$.
% Harmonic Networks: Deep Translation and Rotation Equivariance

\begin{figure*}[!tbp]
\centering
\includegraphics[width=0.7\textwidth]{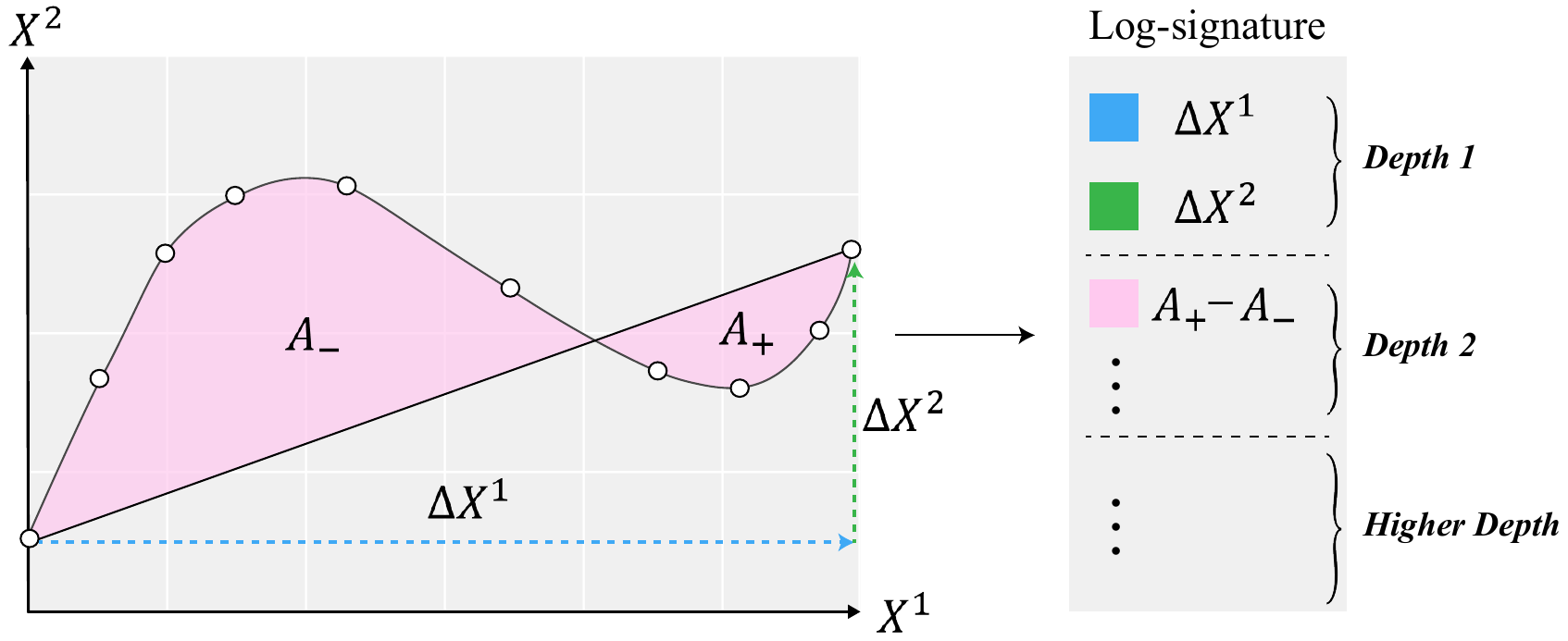}
\caption{Geometric intuition for the first two levels of the log-signature on a 2-dimensional path. We can observe that the depth-1 terms represent the change in each of the coordinates over the interval, and the depth-2 term corresponds to the $L\acute{e}vy~area$ of the path, shown as the signed area enclosed by the curve and the chord connecting its start and endpoints.}
\label{fig:sig_geometry}
\end{figure*}

After that, we turn to analyze the effect of rotation on features acquired by signature transform. Unfortunately, not all elements of a path signature feature exhibit rotation invariance. As demonstrated by~\cite{Diehl2013arXiv}, rotation invariants only exist on levels of even order. For a 2-dimensional continuous path $X:[a,b]\rightarrow\RR^2$ as presented in Fig.~\ref{fig:sig_geometry}, the depth-1 terms corresponding to the variations for each channel over the interval are represented by $\Delta X_1$ and $\Delta X_2$, while the depth-2 term corresponding to the signed area between the chord connecting the endpoints and the real path is denoted as $A$. The calculation of depth-2 term corresponds to the coefficient of the polynomial $[e_n, e_m]$ in Eq.~(\ref{eq:log_sig})
\begin{equation*}
    A = A_+ - A_- = \frac{1}{2}\left(S^{1,2}_{a,b}(X) - S^{2,1}_{a,b}(X)\right).
\end{equation*}
Here, the sign of area exactly indicates the orientation of acceleration, while an increase in the absolute value of surrounding accelerations would increase the proportion of the represented area accordingly. Obliviously, the area $A$ is rotation invariant, while $\Delta X_1$, $\Delta X_2$ do not hold this property. As suggested by~\cite{Diehl2013arXiv}, a depth-2 term in Eq.~(\ref{eq:log_sig}) is rotation invariant only if it involves the iterated integrals over different channels. Although further investigation into rotation invariants for path signature with higher depth over beyond two dimensional paths is still lacking, current conclusion on rotation invariance acquired for depth-2 log-signauture can readily extend to higher dimensional paths.

In our experimental setup, given the coarse grained dynamics $\tilde{\mathbf{X}}_{a,b}\in\RR^{T\times 3M}$, the dimension of depth-$D$ log-signature is
\begin{equation*}
    \mathrm{dim}(\mathrm{LogSig}^D_{a,b}(\tilde{\mathbf{X}})) = \sum_{d=1}^D\frac{1}{d}\sum_{i|d}\mu\left(\frac{d}{i}\right) (3M)^i,
\end{equation*}
where $\mu$ is the M$\ddot{\mathrm{o}}$bius function defined as 
\begin{equation*}
    \mu(n) = \left\{ \begin{array}{rcl}
&0 & \mbox{if $n$ has one or more repeated prime factors} \\
&1 & \mbox{if $n=1$} \\
&(-1)^k & \mbox{if $n$ is the product of $k$ distinct prime numbers}
\end{array}\right.
\end{equation*}
% [Appendix] Neural Rough Differential Equations for Long Time Series 
% [Theorem A.6] Learning stochastic differential equations using RNN with log signature features
When we specify $d=2$, the ratio of rotation-invariant elements over the whole log-signature feature can then be calculated as 
\begin{equation*}
   \gamma  = \frac{3M-1}{3M+1},
\end{equation*}
which indicates the majority of features are rotation-invariant for a large-scale molecular system as we focus on in this paper.
% 能否加实验验证的方式来补充这部分?

% \textbf{General 3D rotations:} can be obtained from three elemental rotations using matrix multiplication:
% \begin{equation}
% \begin{aligned}
% R = R_{z}(\alpha )\,R_{y}(\beta )\,R_{x}(\gamma )
% &={\overset {\text{yaw}}{\begin{bmatrix}\cos \alpha &-\sin \alpha &0\\\sin \alpha &\cos \alpha &0\\0&0&1\\\end{bmatrix}}}{\overset {\text{pitch}}{\begin{bmatrix}\cos \beta &0&\sin \beta \\0&1&0\\-\sin \beta &0&\cos \beta \\\end{bmatrix}}}{\overset {\text{roll}}{\begin{bmatrix}1&0&0\\0&\cos \gamma &-\sin \gamma \\0&\sin \gamma &\cos \gamma \\\end{bmatrix}}}\\ 
% &= {\begin{bmatrix}\cos \alpha \cos \beta &\cos \alpha \sin \beta \sin \gamma -\sin \alpha \cos \gamma &\cos \alpha \sin \beta \cos \gamma +\sin \alpha \sin \gamma \\\sin \alpha \cos \beta &\sin \alpha \sin \beta \sin \gamma +\cos \alpha \cos \gamma &\sin \alpha \sin \beta \cos \gamma -\cos \alpha \sin \gamma \\-\sin \beta &\cos \beta \sin \gamma &\cos \beta \cos \gamma \\\end{bmatrix}}
% \end{aligned}
% \end{equation}

% -----------------------------------------------------------------------
\subsection{Permutation Equivariance}
For the deep clustering module implemented by GNNs $g_{\mathrm{GNN}}(\cdot)$ with the formula presented in Eq.~(\ref{eq:gcn}) as follows
\begin{equation*}
    \mathbf{H}^{l} = \sigma(\tilde{\mathbf{D}}^{-1/2}\tilde{\mathbf{A}}\tilde{\mathbf{D}}^{-1/2}\mathbf{H}^{l-1} \mathbf{W}^{l-1}_{\mathrm{GNN}}).
\end{equation*}
For one layer, the calculation of node embeddings per node involves sum over contributions from different atoms. As a result, node embedding matrix $\mathbf{H}^{l}$ naturally exhibit equivariance with respect to the permutation symmetries of graphs. In addition, the linear mapping shown in Eq.~(\ref{eq:coarse_grain}) only works for dimension alignment, making no influence on the permutation equivariance. So the coarse grained dynamics $\tilde{\mathbf{X}}_{1:T}$ is permutation equivariant with respect to an input set of atoms. 

We then delve into log-signature transform by examining Eq.~(\ref{eq:log_sig}). Notably, the monomials such as $e_n$ and $[e_n, e_m]$ constitute the basis vectors of the vector space, and their coefficients are calculated using iterate integrals over paths indexed by these monomials. Ideally speaking, log-signature features hold permutation invariance since we can arbitrarily select and order the monomials. Such phenomenon becomes even clearer when we revisit the signature transform in Eq.~(\ref{eq:signature}). In the signature transform, the superscripts indicating the paths traverse the set of all multi-indexes denoted as
\begin{equation*}
    W = \{(j_1,\dots,j_d) | d\geq 1,j_1,\dots, j_d\in\{1,\dots,3M\}\}.
\end{equation*}
Hence, the order of the multi-indexes has no impact on the captured dynamic information. In practical implementations, we organize the multi-indexes in ascending order to collect these iterated integral terms~\citep{Kidger2021ICLR}. Consequently, a permutation on the indices of input atoms would yield a predictable permutation transform on the indices of log-signature terms. As a result, we drive the conclusion that our Deep signature method exhibits permutation equivariance with respect to input atom indices.

% -----------------------------------------------------------------------
\subsection{Time-reparametrization invariance}
Since coarse graining will maintain temporal consistency, we only need to analyze the invariance under time-reparametrization for coarse grained dynamics. A reparametrization of a coarse grained pathway $\mathbf{X}:[a,b]\rightarrow\mathbb{R}^{3M}$ is a path $\breve{\mathbf{X}}:[a,b]\rightarrow\mathbb{R}^{3M}$ where $\breve{\mathbf{X}}_t=\mathbf{X}_{\psi_t}$ where $\psi$ is a subjective, continuous, non-decreasing function $\psi:[a,b]\rightarrow[a,b]$. We have the following theorem
\begin{theorem}
    Let $\breve{\mathbf{X}}_t$ be a reparametrization of $\mathbf{X}$, then we have $\mathrm{Sig}(\breve{\mathbf{X}})=\mathrm{Sig}(\mathbf{X})$.
\end{theorem}

% \begin{proof}
Consider two real-valued paths $X,Y:[a,b]\rightarrow\mathbb{R}$ and a surjective, continuous, non-decreasing reparametrization function over time $\psi:[a,b]\rightarrow[a,b]$. Then we have reparametrized paths $\breve{X},\breve{Y}:[a,b]\rightarrow\mathbb{R}$ by $\breve{X}_t=X_{\psi_t}$ and $\breve{Y}_t=Y_{\psi_t}$
\begin{equation*}
    \int_1^T \breve{Y}_t d\breve{X}_t = \int_1^T Y_{\psi_t} d X_{\psi_t} = \int_1^T Y_{\psi_t} \dot{X}_{\psi_t} \dot{\psi_t} dt = \int_1^T Y_{\psi_t} \dot{X}_{\psi_t} d\psi_t
\end{equation*}
After replacing $u=\psi_t$, we have $\int_1^T \breve{Y}_t d\breve{X}_t=\int_1^T Y_u dX_u$, which means path integrals are invariant under a time reparametrization of both paths.  

\noindent Since every term of the signature $S(\mathbf{X})^{j_1,\dots,j_d}_{a,b}$ is defined as an iterated path integral of $\mathbf{X}$, it follows from the above that
\begin{equation*}
    S(\breve{\mathbf{X}})^{j_1,\dots,j_d}_{a,b} = S(\mathbf{X})^{j_1,\dots,j_d}_{a,b}, \quad \forall k\geq 0,~j_1,\dots,j_d \in \{1,\dots,3M\}
\end{equation*}
This complements the proof.
% \end{proof}

% =======================================================================
\section{Experimental Settings}
\label{app:exp_setup}

% -----------------------------------------------------------------------
\subsection{Gene regulatory dynamics}
The dynamics for gene regulatory networks are governed by Michaelis-Menten equation as follows,
\begin{equation}
\label{eq:gene}
    \frac{\mathrm{d} \bm{x}_t(v_i)}{\mathrm{d} t} = -b_i \bm{x}(v_i)^{f} + \sum_{j=1}^n \mathbf{A}^{(i,j)} \frac{\bm{x}^h(v_j)}{\bm{x}^h(v_j) + 1},
\end{equation}
where the first term models the degradation when $f=1$ or dimerization when $f=2$, and the second term represents genetic activation, with the Hill coefficient $h$ determining the level of cooperation in the regulation of the gene.

\noindent\textbf{Data generation.}~For the gene regulatory dynamics, we curate a dataset consisting of 100 trajectories that delineate the intricate interactive dynamics between genes and transcription factors. These trajectories are divided into two distinct categories: degradation type ($f=1$) and dimerization type ($f=2$), each encompassing an equal number of trajectories. To commence a simulation, we first initialize a graph network featuring 100 nodes using a Power-law network generator to elucidate the structural interconnections between these nodes. Subsequently, we employ the Dormand-Prince method to numerically solve the gene regulatory system described by Eq.~(\ref{eq:gene}), with a simulation duration of 2 seconds and a time interval of 0.004 seconds. This computation simulation yields trajectories comprising 500 frames each, laying the foundation for our subsequent experiments.

\noindent\textbf{Model architecture.}~We implement the deep spectral clustering module in our approach using two graph pooling layers, which first coarsen the dynamics into 60 nodes and subsequently into 30 nodes. The dimension of hidden states for the stacked GCN layers is kept as 10. For a tiny implementation, we simply extract the log-signature features for the coarse grained dynamics and then employ a two-layer MLP for property prediction.

\noindent\textbf{Training details.}~We train our model with mini-batches of size 48 for 100 epochs using Adam with the initial learning rate of 5e-4 and a weight decay 1e-4. The coefficients of loss terms are set as $\lambda_1=1$, $\lambda_2=0.01$, and $\lambda_3=10$.

% -----------------------------------------------------------------------
\subsection{Epidermal growth factor receptor mutation dynamics}

\noindent\textbf{Data generation.}~In this investigation, we delve into the intricate binding dynamics between EGFR mutations and RTK, exploring how their interactions can trigger mechanisms of drug resistance. The study encompasses the amalgamation of four RTK partners and five distinct mutation types, alongside the wild type serving as a reference point. To capture the dynamic essence of these interactions, we adhere to the methodological framework outlined in~\citep{Zhu2021BMC}, wherein each system undergoes a meticulous pre-simulation process. This involves an initial energy minimization phase, followed by a 100 picosecond heating stage, subsequent density equilibration spanning 100 picoseconds, and a further 5 nanoseconds of constant pressure equilibration. The equilibrated structures then undergo a simulation period of 50 nanoseconds, resulting in the generation of 24 trajectories. Each trajectory comprises 1000 frames, detailing the temporal interplay among approximately 5,000 atoms. Notably, these trajectories are categorized based on their sensitivity or resistance to the administered drug. For the MD simulations, we leverage the explicit-solvent model integrated within the Amber software suite, utilizing the \emph{Ff99SB} and \emph{gaff} force fields to drive the simulations.

\noindent\textbf{Model architecture.}~For the deep spectral clustering module, we utilize three graph pooling layers that progressively coarsen the dynamics into 400, 200, 50 nodes. The dimension of hidden states for the stacked GCN layers is kept as 20. For the path signature transform module, we partition the input coarse grained dynamics into four segments for the subsequent application of the log-signature transform.

\noindent\textbf{Training details.}~The model is trained with mini-batches of size 16 for a total of 200 epochs. We employ the Adam optimizer with an initial learning rate of 5e-5 and a weight decay 1e-5 to optimize the model parameters. Additionally, the scaling parameters are specified as $\lambda_1=1$, $\lambda_2=0.01$, and $\lambda_3=10$.

% \noindent\textbf{Model architecture.}~We implement the deep spectral clustering module in our approach using two graph pooling layers that first coarsen the dynamics into 60 nodes and then into 30 nodes. The coefficients of loss terms are set as $\lambda_1=1$, $\lambda_2=0.01$, and $\lambda_3=10$.

% We employ the explicit-solvent model built-in Amber software, along with \emph{Ff99SB} and \emph{gaff} force fields to conduct the MD simulation.

% -----------------------------------------------------------------------
\subsection{G protein-coupled receptors dynamics}

\noindent\textbf{Data generation.}~We download our data from the GPCRmd (http://gpcrmd.org/)~\citep{Rodriguez2020GPCR} database, which is an online platform that incorporates web-based visualization capabilities and shares data. This database includes at least one representative structure from each of the four structurally characterized GPCR classes, and holds more than 600 GPCR MD simulations from GPCRmd community and individual contributions. To create our dataset, we select 26 trajectories of the $\beta$2AR-rh1 GPCR inactive (2RH1) and active (3P0G) receptor state with a full agonist. The receptor consists of 282 and 285 amino acids for inactive and active state respectively. Each simulation is conducted for 500 ns with a time interval of 200 ps, therefore every trajectory consists of 2,500 frames that describe the atomic 3D positions over time.

\noindent\textbf{Model architecture.}~We implement the deep spectral clustering module in our approach using two graph pooling layers, which first coarsen the dynamics into 100 nodes and subsequently into 50 nodes. The dimension of hidden states for the stacked GCN layers is kept as 20. In the path signature transform module, the input coarse grained dynamics are partitioned into four segments to facilitate the subsequent application of the log-signature transform.

\noindent\textbf{Training details.}~The model is trained with mini-batches of size 16 for a total of 200 epochs. We employ the Adam optimizer with an initial learning rate of 5e-5 and a weight decay 1e-5 to optimize the model parameters. Furthermore, the scaling parameters are specified as $\lambda_1=1$, $\lambda_2=0.01$, and $\lambda_3=10$.

% =======================================================================
\section{Further Results and Analysis}
\label{sec:app_exp_result}

% -----------------------------------------------------------------------
\subsection{Parameter Sensitivity Analysis}
% \noindent\textbf{The value of $\lambda_1$, $\lambda_3$ and $\lambda_3$.}~
% \begin{table*}[ht]
% \caption{The effect of environmental parameter dimension on the LV dataset. We report the RMSE $(\times \smash{10^{-2})}$ results.}
% \vspace{-0.4cm}
% \label{tab:app_exp_lambda_2}
% % \fontsize{8}{9}\selectfont
% \begin{center}
% \begin{small}
%     \begin{tabular}{cccccc}
%     \toprule[0.7pt]
%     \rule{0pt}{2ex}  & 10 & 1 & 0.1 & 0.01  \\
%     \hline  
%     \rule{0pt}{2ex} $\lambda_1$    &65.27\scalebox{0.65}{$\pm2.48$} &52.67\scalebox{0.65}{$\pm0.94$} &3.736  &5.659 \\
%     \rule{0pt}{2ex} $\lambda_2$     &34.751 &42.632 &33.774  &50.981  \\ 
%     \bottomrule[0.7pt]
%     \end{tabular}
% \end{small}
% \end{center}
% \vspace{-0.2cm}
% \end{table*}

\noindent\textbf{The sensitivity of coarsening level.}~To further validate the robustness of our method for different coarsening levels, we conducted additional experiments by varying the coarsening levels for the EGFR and GPCR datasets. The results are summarized in Table~\ref{tab:coarsening_accuracy}.
\begin{table*}[h!]
\caption{Comparison of EGFR and GPCR Accuracy at Different Coarsening Levels}
\label{tab:coarsening_accuracy}
\vspace{-0.2cm}
\centering
\begin{small}
    \begin{tabular}{cccccc}
    \toprule[0.7pt]
    \textbf{Coarsening Level} & \textbf{100} & \textbf{50} & \textbf{30} & \textbf{10} & \textbf{5} \\ 
    \hline 
    \textbf{EGFR Accuracy (\%)} & 66.93\scalebox{0.65}{$\pm4.62$} & 69.33\scalebox{0.65}{$\pm4.78$} & 66.40\scalebox{0.65}{$\pm5.06$} & 63.20\scalebox{0.65}{$\pm7.56$} & 57.33\scalebox{0.65}{$\pm4.73$} \\ 
    \textbf{GPCR Accuracy (\%)} & 50.67\scalebox{0.65}{$\pm5.58$} & 58.00\scalebox{0.65}{$\pm4.17$} & 58.00\scalebox{0.65}{$\pm6.01$} & 57.73\scalebox{0.65}{$\pm8.92$} & 55.07\scalebox{0.65}{$\pm6.32$} \\
    \bottomrule[0.7pt]
    \end{tabular}
\end{small}
\end{table*}

From these results, we observe that coarsening levels of 30 and 50 beads yield the best accuracy for both datasets, demonstrating the robustness of our method across different levels of coarsening. While extreme levels, such as 100 or 5 beads, result in reduced performance, the method is flexible and performs consistently well within a broad range of coarsening levels.

\noindent\textbf{The value of $\lambda_1$, $\lambda_2$ and $\lambda_3$.}~In the objective of Deep Signature presented in Eq.~(\ref{eq:loss_total}), there exist three hyper-parameters $\lambda_1$, $\lambda_2$ and $\lambda_3$ reweighting the loss terms. We now conduct the parameter sensitivity test by varying their value in $\{10.0, 1.0, 0.1, 0.01\}$. The results are summarized in Table~\ref{tab:lambda1}, \ref{tab:lambda2} and \ref{tab:lambda3}, respectively.
\begin{table*}[h!]
\caption{Sensitivity analysis for $\lambda_1$ and its effect on accuracy.}
\label{tab:lambda1}
\vspace{-0.2cm}
\centering
\begin{tabular}{ccccc}
\toprule[0.7pt]
\textbf{$\lambda_1$}       & \textbf{10.0} & \textbf{1.0} & \textbf{0.1} & \textbf{0.01} \\ 
\hline
\textbf{Accuracy (\%)}    & 59.60\scalebox{0.65}{$\pm8.34$}  & 69.33\scalebox{0.65}{$\pm4.78$} & 67.87\scalebox{0.65}{$\pm1.95$} & 67.33\scalebox{0.65}{$\pm2.77$}  \\ 
\bottomrule[0.7pt]
\end{tabular}
\end{table*}

\begin{table*}[h!]
\caption{Sensitivity analysis for $\lambda_2$ and its effect on accuracy.}
\label{tab:lambda2}
\vspace{-0.2cm}
\centering
\begin{tabular}{ccccc}
\toprule[0.7pt]
\textbf{$\lambda_2$}       & \textbf{10.0} & \textbf{1.0} & \textbf{0.1} & \textbf{0.01} \\ 
\hline
\textbf{Accuracy (\%)}    & 65.27\scalebox{0.65}{$\pm2.48$}  & 67.67\scalebox{0.65}{$\pm1.63$}  & 67.80\scalebox{0.65}{$\pm1.15$} & 69.33\scalebox{0.65}{$\pm4.78$}  \\
\bottomrule[0.7pt]
\end{tabular}
\end{table*}

\begin{table*}[h!]
\caption{Sensitivity analysis for $\lambda_3$ and its effect on accuracy.}
\label{tab:lambda3}
\vspace{-0.2cm}
\centering
\begin{tabular}{ccccc}
\toprule[0.7pt]
\textbf{$\lambda_3$}       & \textbf{10.0} & \textbf{1.0} & \textbf{0.1} & \textbf{0.01} \\ \hline
\textbf{Accuracy (\%)}    & 69.33\scalebox{0.65}{$\pm4.78$}  & 66.53\scalebox{0.65}{$\pm2.65$}  & 62.00\scalebox{0.65}{$\pm4.22$} & 58.93\scalebox{0.65}{$\pm1.71$}  \\ 
\bottomrule[0.7pt]
\end{tabular}
\end{table*}

As shown, our model is generally robust to moderate variations in the hyperparameters. For instance, $\lambda_1 = 1.0$, $\lambda_2 = 0.01$, and $\lambda_3 = 10.0$ consistently deliver strong performance. However, extreme values for these parameters (\eg~$\lambda_3 = 0.01$ or $\lambda_1 = 10.0$) can lead to a noticeable drop in accuracy, highlighting the importance of choosing balanced values.

\end{document}